\documentclass[11pt,a4]{article}
\usepackage{amsmath,amsfonts,setspace,url,hyperref,verbatim,cite}
\usepackage[utf8]{inputenc}
\usepackage{fullpage}
\usepackage{breqn}
\newcommand{\be}{\begin{equation}}
\newcommand{\ee}{\end{equation}}

\newcommand{\cH}{\mathcal{H}}

\newcommand{\Xx}{\mathbb{X}}
\newcommand{\utau}{\underline\tau}
\newcommand{\uu}{\underline u}
\newcommand{\um}{\underline m}
\newcommand{\un}{\underline n}
\newcommand{\up}{\underline p}

\newcommand{\ba}{\bar{a}}
\newcommand{\bb}{\bar{b}}

\newcommand{\Pp}{P}
\newcommand{\Pm}{\bar{P}}
\newcommand{\Vp}{V}
\newcommand{\Vm}{\bar{V}} 

\newcommand{\pA}{A}
\newcommand{\pB}{B}
\newcommand{\pC}{C}
\newcommand{\pD}{D}
\newcommand{\pE}{E}
\newcommand{\pF}{F}

\newcommand{\mA}{\bar{A}}
\newcommand{\mB}{\bar{B}}
\newcommand{\mC}{\bar{C}}
\newcommand{\mD}{\bar{D}}
\newcommand{\mE}{\bar{E}}
\newcommand{\mF}{\bar{F}}

\newcommand{\flathm}{\boldsymbol{\bar{h}}}
\newcommand{\flathp}{\boldsymbol{h}}

\newcommand{\bx}{\bar{x}}
\newcommand{\by}{\bar{y}}
\newcommand{\hv}{\hat{v}}
\newcommand{\bh}{\bar h} 

\newcommand{\uH}{\mathring{\mathcal{H}}}
\newcommand{\uf}{\mathring{f}}

\numberwithin{equation}{section}

\usepackage{tikz}
\usetikzlibrary{matrix,arrows,positioning,shapes,snakes,fadings,decorations.pathmorphing,
decorations.pathreplacing,automata,shadows,decorations.markings,patterns}
\usepackage[thicklines]{cancel}
\usetikzlibrary{calc,shapes.callouts,shapes.arrows}
\newcommand*\circled[1]{\footnotesize\tikz[baseline=(char.base)]{%
            \node[thick,shape=circle,fill=blue!5,draw,inner sep=2pt] (char) {#1};}}
            \usepackage{enumitem}

\begin{document}

\title{
\bf
A worldsheet supersymmetric Newton-Cartan string}

\author{\sc Chris D. A. Blair\footnote{\tt cblair@vub.ac.be}}

\date{
Theoretische Natuurkunde, Vrije Universiteit Brussel, and the International Solvay Institutes,  Pleinlaan 2, B-1050 Brussels, Belgium 
}

\maketitle

\begin{abstract}
We construct a (locally) supersymmetric worldsheet action for a string in a non-relativistic Newton-Cartan background.
We do this using a doubled string action, which describes the target space geometry in an $O(D,D)$ covariant manner using a doubled metric and doubled vielbeins.
By adopting different parametrisations of these doubled background fields, we can describe both relativistic and non-relativistic geometries.
We focus on the torsional Newton-Cartan geometry which can be obtained by null duality/reduction (such null duality is particularly simple for us to implement). 
The doubled action we use gives the Hamiltonian form of the supersymmetric Newton-Cartan string action automatically, from which we then obtain the equivalent Lagrangian.
We extract geometric quantities of interest from the worldsheet couplings and write down the supersymmetry transformations. 
Our general results should apply to other non-relativistic backgrounds. We comment on the usefulness of the doubled approach as a tool for studying non-relativistic string theory.
\end{abstract}

\tableofcontents

\section{Introduction}

This is a paper about supersymmetric non-relativistic string theory, and it exists because the author was surprised.

The cause of the surprise was a connection between two very different sounding topics.  The first is the description of strings in non-relativistic Newton-Cartan backgrounds, which has recently been explored extensively in (for instance) \cite{Harmark:2017rpg, Kluson:2018uss,
Bergshoeff:2018yvt, 
Hansen:2018ofj, Harmark:2018cdl,Bergshoeff:2018vfn,Kluson:2018egd,Kluson:2018grx,Gomis:2019zyu,Kluson:2019qgj,Gallegos:2019icg,Harmark:2019upf,Bergshoeff:2019pij,Roychowdhury:2019olt,Roychowdhury:2019vzh, Roychowdhury:2019qmp}.
This is in part inspired by motivations from holography \cite{Christensen:2013lma,Christensen:2013rfa}, but also recalls older studies of non-relativistic limits of string theory \cite{Gomis:2000bd, Danielsson:2000gi,  Danielsson:2000mu}, in which one might hope to find a novel corner of string theory in which at least some aspects of the full theory become simpler to understand.

The second is the ``doubled'' approach to manifest T-duality covariance in string theory or supergravity.
Here the basic idea is to extend the geometry of spacetime in a way that leads to an immediately $O(D,D)$ covariant theory. 
These doubled approaches include doubled worldsheet actions such as \cite{Tseytlin:1990nb, Tseytlin:1990va, Hull:2004in, Hull:2006va} (in which we introduce twice the number of target space coordinates, plus a chirality constraint to ensure the number of degrees of freedom on the worldsheet remains the same) and the double field theory approach to supergravity \cite{Siegel:1993th,Siegel:1993xq,Hull:2009mi} (in which we formally introduce dual coordinates on which the spacetime fields may in principle depend, along with an $O(D,D)$ covariant constraint which restricts to the usual number of coordinates).

Remarkably, it was realised in \cite{Lee:2013hma, Ko:2015rha,Morand:2017fnv} that non-relativistic geometries such as Newton-Cartan have a home in these doubled approaches.
This is surprising because this does not seem like something you would naturally expect to find in a formulation intended to describe T-duality in a relativistic theory. 
Surprise, of course, is a function of ignorance. An explanation for the surprise is that the worldsheet description of strings in certain Newton-Cartan backgrounds can be related to strings in relativistic backgrounds with a null isometry using a sort of T-duality transformation \cite{Harmark:2017rpg, Bergshoeff:2018yvt,Harmark:2018cdl}. 
From the relativistic side, this appears to be ill-defined (recall the Buscher rule inverts the metric component in the isometry direction, $\tilde g_{zz} = \frac{1}{g_{zz}}$, but $g_{zz} = 0$ if $z$ is a null isometry), so has to be interpreted carefully.
In the doubled approach, however, we choose to work with $O(D,D)$ valued background fields rather than the usual spacetime metric and $B$-field, and even if the latter are seemingly nonsensical the former need not be. Then this sort of null duality is formally well-defined, and issues only arise when trying to use an inappropriate spacetime parametrisation of the $O(D,D)$ valued fields. The \emph{appropriate} parametrisation in fact \cite{Morand:2017fnv} will turn out to describe a non-relativistic geometry! (The surprise is then that the same doubled formalism admits both relativistic and non-relativistic parametrisation in the first place. 
This surprise can also be uplifted to M-theory: for an initial exploration of these ideas in the U-duality covariant ``exceptional'' formalism, see \cite{Berman:2019izh}.)

In this paper, we will explicitly connect the dots between a particular worldsheet supersymmetric doubled sigma model \cite{Blair:2013noa} and the action for a worldsheet supersymmetric Newton-Cartan string, extending the Polyakov action of \cite{Harmark:2018cdl}. 

The action of \cite{Blair:2013noa} was motivated by the natural appearance of $O(D,D)$ covariant structures in the Hamiltonian approach to worldsheet string theory.
Note that the Hamiltonian analysis of the Newton-Cartan string, and the use of null T-duality in this setting, has been explored in \cite{Kluson:2018uss,Kluson:2018grx,Kluson:2018egd,Kluson:2019qgj} (and extended to other branes in non-relativistic backgrounds \cite{Kluson:2019ifd,Kluson:2019uza,Kluson:2019avy}). Effectively, what will help here is that limits which appear singular in the Lagrangian description may be non-singular in the corresponding Hamiltonian picture. 
By working not with the ordinary spacetime metric and form fields as fundamental fields but instead with a larger generalised metric, we can achieve similar results.

This also means that at the level of the worldsheet, the doubled string sigma models that we will use below can be effectively viewed as the standard string action in Hamiltonian form, where we have defined new worldsheet scalars $\tilde X$ in terms of the canonical momenta $P$ by $\tilde X^\prime = P$. The background fields appear in parametrisations of an $O(D,D)$-valued generalised metric, $\mathcal{H}_{MN} ( g, B)$ (and its corresponding vielbein). If we allow ourselves within the Hamiltonian framework to range over all possible parametrisations of $\mathcal{H}_{MN}$ including those that are not consistent with having a standard relativistic metric, then we will discover non-relativistic geometries and other more exotic ``non-Riemannian'' scenarios.
There is not necessarily any need to then invoke the tools and interpretation of the doubled formalism; however the latter will be especially useful in studying features of the spacetime theory as its kinematics and dynamics are known irrespective of parametrisation.

Bosonic particle and string actions in non-relativistic or non-Riemannian backgrounds have been obtained from the doubled formalism already in the papers \cite{Lee:2013hma, Ko:2015rha,Morand:2017fnv}
The action we construct will be supersymmetric on the worldsheet (i.e. it is a spinning or RNS string). 
A doubled Green-Schwarz action has also been constructed in \cite{Park:2016sbw}, in which the application to the Gomis-Ooguri non-relativistic string was considered.

\subsection*{Outline and main result}

The principal points of this paper are as follows.

\begin{enumerate}[label=\protect\circled{\arabic*}]
\item We first demonstrate explicitly in the bosonic setting how a doubled action can handle a null duality leading to the Newton-Cartan Polyakov action of \cite{Harmark:2018cdl}. This is the subject of section \ref{bosonic}. 
\item We discuss some general features of the effect of null duality on the transformation rules of vielbeins and worldsheet fermions, in section \ref{vielbeininterlude}. We point out that the spacetime vielbein $e^A{}_\mu$ transforms to different \emph{non-invertible} spacetime vielbeins $e_\pm^A{}_\mu$ in the left- and right-moving sectors (which the left- and right-moving fermions should couple to), and that we cannot find a spacetime Lorentz transformation that relates these two quantities. This underlies the fact that we obtain a non-relativistic geometry: the left- and right-moving sectors of the string no longer see the same relativistic target space. The counterpart of this in the bosonic sector is that a certain pair of directions in the target space become chiral and antichiral respectively. However, we do not lose any degrees of freedom once we take into account that these chirality constraints are imposed by an extra pair of fields appearing as Lagrange multipliers.

Despite these issues, we observe that the doubled vielbeins appearing in the doubled string are perfectly well-defined after the duality. This helps us understand what we should use as an appropriate basis for the worldsheet fermions in the Newton-Cartan string.

\item In section \ref{NCRNS}, after reviewing the worldsheet supersymmetric doubled string of \cite{Blair:2013noa}, we show that the Polyakov action for a string in a Newton-Cartan background has the following worldsheet supersymmetric extension:
\[
\begin{split}
S = \int d^2\sigma\, 
& 
	\frac{1}{2} h_{ij} \left( \frac{1}{e}D_\tau X^i D_\tau X^j  - e X^{\prime i} X^{\prime j} \right) 
	+ B_{\mu\nu}  \dot X^{ \mu} X^{\prime \nu} 
\\ &
- \frac{i}{2} \left( \psi^{\mA}  \flathm_{\mA \mB} D_+ \psi^{\mB} 
+  \psi^{\mA} D_+ X^i \omega_{+ i \mA \mB} \psi^{\mB}
\right)
\\ & 
- \frac{i}{2}\left(
 \tilde\psi^{\pA}  \flathp_{\pA \pB} D_- \tilde\psi^{\pB} 
+ \tilde\psi^{\pA} D_- X^i \omega_{- i \pA \pB} \tilde\psi^{\pB}
\right)
\\ & 
- \frac{i}{2e} \tilde \xi ( e_{i \um} \tilde \psi^{\um} + \tau_i \tilde\psi^{\uu}  ) D_+ X^i
+ \frac{i}{2e}  \xi ( e_{i \um}   \psi^{\um} + \tau_i \psi^{\uu} ) D_- X^i
\\ & 
- \frac{1}{12} T_{\pA \pB \pC} \tilde \xi \tilde\psi^{\pA} \tilde\psi^{\pB} \tilde\psi^{\pC} 
- \frac{1}{12} \bar T_{\mA \mB \mC}  \xi \psi^{\mA} \psi^{\mB} \psi^{\mC} 
- \frac{1}{4e} \tilde \xi \xi \flathp_{\pA \mB} \tilde \psi^{\pA} \psi^{\mB}
\\ 
& + \frac{e}{2} R_{\mA \mB \pC \pD} \psi^{\mA} \psi^{\mB} \tilde\psi^{\pC} \tilde\psi^{\pD} 
\\
& + 
 \boldsymbol{\beta}
\left(
D_- V + \tau_i D_- X^i - ie \sqrt{2} \Phi_{\uu \mA \mB} \psi^{\mA} \psi^{\mB}
- i \tilde \xi \tilde \psi^{\utau}  \right)
\\ &
- 
\boldsymbol{\bar\beta}
\left(
D_+ V -\tau_i D_+ X^i + ie \sqrt{2} \tilde\Phi_{\uu \pA \pB} \tilde\psi^{\pA} \tilde \psi^{\pB}
- i\xi \psi^{\utau}
\right)
 \,.
\end{split}
\]
Here $i=1,\dots,d$ and the Newton-Cartan geometry is described by the pair $(h_{ij},\tau_i)$, where $h_{ij}$ is symmetric and has rank $d-1$, with zero vector $v^i$ such that $h_{ij}v^j = 0$; the covector $\tau_i$ specifies the preferred Newton-Cartan time direction with $v^i \tau_i = -1$.
The index $\um$ is a flat index with $\um =1,\dots ,d-1$ and we have a pseudo-vielbein $e_{i}{}^{\um}$ such that $h_{ij} = e_i{}^{\um} e_j{}^{\un} \delta_{\um\un}$.
The bosonic target space coordinates are $X^\mu = (X^i,V)$.
If we view this action as being obtained from null duality, then $V$ is the coordinate dual to the original null isometry direction.
We have additional worldsheet fields $\boldsymbol{\beta}$ and $\boldsymbol{\bar\beta}$, which enforce what can be viewed as chirality conditions on $V \pm \tau_i X^i$.
In the Hamiltonian approach, $\boldsymbol{\beta}$ and $ \boldsymbol{\bar\beta}$ arise from components of the momenta conjugate to $X^\mu$ that do not appear quadratically in the action, and cannot be integrated out.

The worldsheet fermions $\psi^{\mA}$ and $\tilde\psi^{\pA}$ are one-component Majorana-Weyl spinors and anticommuting. They carry flat indices $\mA = 1,\dots,d+1$ and $\pA =1,\dots,d+1$ associated to separate chiral $O(1,d)$ groups. Note however that only a common ${O}(d-1)$ subgroup of these can be realised as the conventional background local symmetry group, as a consequence of the non-relativistic parametrisation we will specify. 
We decompose the flat indices such that $\psi^{\mA} = ( \psi^{\um} , \psi^{\utau},\psi^{\uu} )$ and $\tilde\psi^{\pA} = ( \tilde \psi^{\um} , \tilde\psi^{\utau},\tilde \psi^{\uu} )$, where $\um=1,\dots, d-1$. Here the indices $\utau$ and $\uu$ do not run over anything, and label the fermions which are the superpartners of $(\tau_i X^i \pm V)$ and $\boldsymbol{\beta}, \boldsymbol{\bar \beta}$ respectively (see below). (The notation is explained in section \ref{NCvieltech}.) 
These indices are contracted using flat metrics, $\flathm_{\mA\mB}$ and $\flathp_{\pA \pB}$, with $\flathm_{\um\un} = \flathp_{\um\un} =\delta_{\um\un}$ and $\flathm_{\utau\uu} = \flathp_{\utau\uu} =1$.

The $B$-field includes the additional Newton-Cartan $U(1)$ gauge field $m_i$ as the component $B_{iv} = -m_i$. We will also consider components $B_{ij} \neq 0$.
Further couplings to the background are contained in the spin connections, $\omega_{+ i \mA \mB}$, $\omega_{-i \pA \pB}$, torsions $T_{\pA \pB \pC}$ and $\bar T_{\mA \mB \mC}$, and curvature $R_{\mA \mB \pC \pD}$, which are defined in \eqref{omegasNC}, \eqref{torsionsNC} and \eqref{curvNC} in terms of certain geometric quantities arising automatically in the doubled approach. These geometric quantities are in effect certain (combinations of) projections of doubled spin connections. The quantities $\Phi_{\uu \mA \mB}$ and $\tilde\Phi_{\uu \pA \pB}$ also appearing in this action are also of this nature. These details will be explained in the course of the paper.

Finally, the worldsheet derivatives are $D_\tau \equiv \partial_\tau - u \partial_\sigma$, $D_\pm \equiv D_\tau \pm e \partial_\sigma$, where $e$ and $u$ are the two independent components of the worldsheet metric. The superpartners of $e$ and $u$ are $\xi$ and $\tilde \xi$, which are one-component anticommuting Majorana-Weyl spinors, and are the independent components of the worldsheet gravitino. Our worldsheet conventions are contained in appendix \ref{ws}, from which one can also check that the form of the action we have written above can be made manifestly covariant on the worldsheet.

\item As well as the action, we write down the supersymmetry transformations in section \ref{susytransfs}. This tells us something about the fermionic counterpart of the ``constraints'' imposed by the equations of motion $\boldsymbol{\beta}$ and $\boldsymbol{\bar\beta}$. This is most intuitive in a flat background: there these constraints enforce that certain combinations of the bosonic coordinates are chiral. Now, the fermions are automatically also chiral, and the bosonic constraints transform into the equations of motion for the superpartners of $\boldsymbol{\beta}, \boldsymbol{\bar\beta}$, which are just the standard fermionic equations of motion for certain combinations of the fermions. We use a basis in which this is automatic: with our pairs of superpartners being $(\boldsymbol{\beta}, \tilde\psi^{\uu})$, $( \boldsymbol{\bar\beta}, \psi^{\uu})$, $(V+\tau_iX^i, \tilde\psi^{\utau})$, $(-V+\tau_iX^i, \psi^{\utau})$.  

\item In addition, in the discussion in section \ref{discussion}, we generate a couple of simple examples of potential Newton-Cartan backgrounds based on null or timelike dualities in the doubled setting, and suggest some advantages (and disadvantages) of using the doubled approach to further understand non-relativistic string theory.

\end{enumerate}

\section{Newton-Cartan string and doubled string} 
\label{bosonic}

\subsection{Newton-Cartan from null duality} 

\subsubsection*{Newton-Cartan variables}

First let us recall from \cite{Berman:2019izh} how to embed the Newton-Cartan geometry of \cite{Harmark:2017rpg, Harmark:2018cdl} into doubled language.
This geometry can be conveniently first viewed in terms of a $d+1$ dimensional Lorentzian spacetime with a null isometry, which can always be put into the form
\be
ds^2 = g_{\mu\nu} dx^\mu dx^\nu = 2 \tau_i dx^i ( du - m_i dx^i ) + h_{ij} dx^i dx^j \,,
\label{metric1}
\ee
where $u$ denotes the null direction, and the $d$ dimensional matrix $h_{ij}$ has rank $d-1$.
The fields $\tau_i, m_i$ and $h_{ij}$ together describe a torsional Newton-Cartan geometry. The objects $(h_{ij}, \tau_i)$ can be viewed as a pair of degenerate metrics, while $m_i$ is a $U(1)$ gauge field associated to mass conservation.
We can also introduce a vector $v^i$ 
and a rank $d-1$ matrix $h^{ij}$ such that 
\be
h_{ij} v^j = 0 \,,\quad v^i \tau_i = -1 \,,\quad
h^{ij} \tau_j = 0 \,,\quad h^{ik} h_{kj} - v^i \tau_j = \delta^i_j \,.
\ee
It is convenient to also define
\be
\bh_{ij} \equiv h_{ij} - \tau_i m_j - \tau_j m_i \,,\quad
\hv^i \equiv v^i - h^{ij} m_{j} \,,\quad
\tilde \Phi \equiv - v^i m_i + \frac{1}{2} h^{ij} m_i m_j \,,
\ee
such that the completeness holds also as $h^{ik} \bh_{kj} - \hv^i \tau_j = \delta^i_j$.
The inverse metric is
\be
g^{\mu\nu} = \begin{pmatrix}
h^{ij} & - \hv^i \\ 
- \hv^j & 2 \tilde \Phi 
\end{pmatrix} \,,
\ee
while the determinant is 
\be
\det g = - \frac{\det \bh}{2\Phi} = \frac{1}{ \frac{1}{(d-1)!} \epsilon_{i_1 \dots i_d} \epsilon_{j_1 \dots j_d} v^{i_1} v^{j_1} h^{i_2 j_2} \dots h^{i_d j_d} }\,.
\ee
We would like to ``dualise'' this model on the null isometric direction. The conventional Buscher rules involve inverting the metric component $g_{uu}$, which is of course zero here.
Despite this, one can indeed carry out this sort of dualisation by introducing a Lagrange multiplier $A_\alpha$ and new worldsheet scalar $V$ in order to place the momenta conjugate to the direction $U$ on-shell \cite{Harmark:2017rpg, Harmark:2018cdl}, leading to an action for a string in a Newton-Cartan background (see also \cite{Bergshoeff:2018yvt}, which showed that a non-relativistic string in the somewhat different so-called stringy Newton-Cartan background is also T-dual to a Lorentzian background with a null isometry - the two actions are related in \cite{Harmark:2019upf}).

\subsubsection*{Doubled variables}

We can reinterpret this procedure by embedding the null duality in the ``doubled'' framework, which encompasses both worldsheet models \cite{Tseytlin:1990nb, Tseytlin:1990va, Hull:2004in, Hull:2006va} and the target space supergravity via double field theory \cite{Siegel:1993th,Siegel:1993xq,Hull:2009mi}.
The conceptual advantage here for us will be the repackaging of the original spacetime (or Newton-Cartan space and time) background into quantities which transform covariantly under general $O(D,D)$ T-duality transformations. This will allow us additional ``freedom'' to evade the singularities that would otherwise appear in the Buscher rules. 

Let us introduce some notation. 
Any object in the fundamental representation of $O(D,D)$ carries a doubled index, $M,N,\dots=1,\dots,2D$, which decomposes into spacetime vector and covector indices such that $V^M = ( V^\mu, V_\mu )$ (for the metric in \eqref{metric1}, we have $D=d+1$).
By definition $O(D,D)$ transformations preserve the following bilinear form:
\be
\eta_{MN} = \begin{pmatrix} 0 & I \\ I & 0 \end{pmatrix} \,,
\ee
which is used together with its inverse $\eta^{MN}$ to raise and lower doubled indices.
We package the NSNS sector fields $(g_{\mu\nu}, B_{\mu\nu}, \phi)$ into a generalised metric, $\cH_{MN}$, and generalised dilaton, $d$. The latter is a scalar under $O(D,D)$ transformations, and would normally be related to the determinant of the spacetime metric and the dilaton $\phi$ by $e^{-2d} = e^{-2\phi} \sqrt{|\det g|}$.
The generalised metric is defined to obey:
\be
\cH_{MN} = \cH_{NM} \,,\quad \cH_{MK} \eta^{KL} \cH_{LN} = \eta_{MN} \,,
\label{gmconds}
\ee
and the standard solution of these constraints is to take $\cH_{MN}$ to parametrise the coset $O(D,D) / O(D) \times O(D)$ or $O(D,D) / O(1,D-1) \times O(1,D-1)$. For the Newton-Cartan application we are interested in, the latter Lorentzian coset is appropriate. Note that other solutions exist \cite{Morand:2017fnv}, which we will see below, which may have uses for other non-relativistic geometries.
The standard parametrisation of the generalised metric in $O(D,D) / O(1,D-1) \times O(1,D-1)$ is
\be
\cH_{MN} = \begin{pmatrix} g_{\mu\nu} - B_{\mu\rho} g^{\rho\sigma} B_{\sigma \nu} & B_{\mu\rho} g^{\rho \nu} \\- g^{\mu\rho} B_{\rho \nu} & g^{\mu\nu} \end{pmatrix}  \,.
\label{standardparam}
\ee

\subsubsection*{Generating the Newton-Cartan generalised metric}

We now will carry out a null duality on the background \eqref{metric1}, viewing this as a particular $O(D,D)$ transformation acting on the generalised metric.
So, first we insert the metric \eqref{metric1} into the standard parametrisation \eqref{standardparam} of the generalised metric.
We assume for now there is no background $B$-field (however we will see at the end of this section that it is straightforward to incorporate one), so that we simply have $\cH_{MN} = \mathrm{diag}\,( g_{\mu\nu}, g^{\mu\nu})$. 
To carry out the analogue of a Buscher transformation on the null isometry direction, we split $\mu = ( i , u)$ and then act on the generalised metric with the $O(D,D)$ transformation which swaps the ${}^u$ and ${}_u$ indices, namely:
\be
\mathcal{T}^M{}_N = \begin{pmatrix}
\delta^i{}_j  & 0 & 0 & 0 \\
0 & 0 & 0 & 1 \\
0 & 0 & \delta_i{}^j & 0 \\
0 & 1 & 0 & 0 
\end{pmatrix} \,.
\ee
(It is worth emphasising that this means that we can also do the inverse problem with no difficulties.)
Let's note first that the invariant generalised dilaton (assuming $\phi=0$ before the duality) is 
\be
e^{-2d} = \sqrt{ \left|\frac{\det \bh}{2\Phi}\right|} = \sqrt{\frac{1}{| \frac{1}{(d-1)!} \epsilon_{i_1 \dots i_d} \epsilon_{j_1 \dots j_d} v^{i_1} v^{j_1} h^{i_2 j_2} \dots h^{i_d j_d}| }}\,. 
\ee
More immediately interesting and useful is the dual generalised metric, which we denote by $\cH_{\text{NC}}$ as we will refer to it as the Newton-Cartan generalised metric:
\be
(\cH_{\text{NC}})_{MN} = \begin{pmatrix} 
\bh_{ij} & 0 & 0 & \tau_i \\ 
0 & 2 \tilde \Phi & - \hv^j & 0 \\
0 & - \hv^i & h^{ij} & 0 \\
\tau_j & 0 & 0 & 0 
\end{pmatrix} \,.
\label{NCGM}
\ee
We see immediately that this does not admit the standard parametrisation of \eqref{standardparam} because the lower right $D \times D$ block is not invertible and so cannot be interpreted as the inverse spacetime metric!

\subsubsection*{General parametrisations of generalised metrics}

However, in a doubled approach, the generalised metric \eqref{NCGM} is a perfectly well-defined object.
Indeed, a classification of all possible parametrisations of the generalised metric subject to the conditions \eqref{gmconds} was carried out in \cite{Morand:2017fnv}.
These parametrisations take the general form
\be
\cH_{MN} = 
\begin{pmatrix}
1 & B \\ 0 & 1 
\end{pmatrix} 
\begin{pmatrix}
K& Z \\
Z^T & H 
\end{pmatrix} 
\begin{pmatrix}
1 & 0 \\ -B & 1 
\end{pmatrix} 
\label{generalgM}
\ee
where the matrices $K_{\mu\nu}$ and $H^{\mu\nu}$ are simultaneously degenerate, each having $n+\bar n$ zero eigenvectors. 
Let a basis for the null eigenvectors of $H^{\mu\nu}$ be $x_\mu^I$, $I=1,\dots,n+\bar n$, and a dual basis for those of $K_{\mu\nu}$ be $y^\mu_I$, with
\be
x_\mu^I y^\mu_{I^\prime} = \delta^I_{I^\prime} \,, \quad
H^{\mu\rho} K_{\rho \nu} + x_\nu^I y_I^\mu = \delta^\mu_\nu \,.
\ee
Then the matrix $Z_\mu{}^\nu = x_\mu^I \sigma_I{}^{I^\prime} y^\nu_{I \prime}$, where the matrix $\sigma_I{}^{I^\prime}$ has eigenvalues $+1$ with multiplicity $n$ and $-1$ with multiplicity $\bar{n}$. A canonical choice of bases then consists of $x_\mu^a, \bx_\mu^{\ba}$, $y^\mu_a$, $\by^\mu_{\ba}$, $a=1,\dots,n$ and $\ba = 1,\dots \bar n$, such that
\be
x_\mu^a y^\mu_b = \delta^a_b \,,\quad \bx_\mu^{\ba} y^\mu_{\bar b} = \delta^{\ba}_{\bar b}\,,\quad
x_\mu^a \by^\mu_{\bar b} = 0 = \bx_\mu^{\ba} y^{\mu}_{  b} \,,
\ee
\be
Z_\mu{}^\nu = x_\mu^a y^\nu_a - \bx_\mu{}^{\ba} \by^\nu_{\ba} 
\,,\quad
H^{\mu\rho} K_{\rho \nu} + x_\nu^a y_a^\mu + \bx_\nu^{\ba} \by_{\ba}^{\mu} = \delta^\mu_\nu \,.
\ee
The integers $n$ and $\bar{n}$ characterise the type of parametrisation, and appear in the trace $\cH^M{}_M = 2(n-\bar{n})$. Evidently, in a usual ``Riemannian'' parametrisation, $n=\bar{n} = 0$. All other cases are then non-Riemannian in nature, as the block $H^{\mu\nu}$ is not invertible and so cannot be interpreted as a spacetime metric. 
This does not mean these other cases are not geometric: they may just be geometries of different type.
As shown in \cite{Morand:2017fnv}, this implies that many versions of non-relativistic geometries, including Newton-Cartan, can be embedded as a generalised metric and hence understood in the doubled formalism. 

A non-Riemannian background with $n = \bar{n} \neq 0$ can be generated by $O(D,D)$ transformations acting on a Riemannian generalised metric. This is what happens when we obtain the Newton-Cartan geometry by starting with the background \eqref{metric1} and dualising on the null duality.
Another closely related example is the T-duality of a fundamental string solution on both the time and string spatial direction, which for particular values of the original $B$-field gives rise to a non-Riemannian background which may be related to the Gomis-Ooguri string \cite{Gomis:2000bd, Danielsson:2000gi,  Danielsson:2000mu}, as shown in \cite{Lee:2013hma,Ko:2015rha}.

One point worth mentioning is that the decomposition of a given non-Riemannian generalised metric into $H^{\mu\nu}$, $K_{\mu\nu}$, $B_{\mu\nu}$ is not unique, owing to the presence of certain shift symmetries \cite{Morand:2017fnv}. In the Newton-Cartan case these will actually correspond to Galilean transformations.

\subsubsection*{Back to Newton-Cartan}

The Newton-Cartan generalised metric \eqref{NCGM} admits a parametrisation of the form \eqref{generalgM} with $n=\bar{n} =1$ and:
\be
K_{\mu\nu} = \begin{pmatrix} h_{ij} & 0 \\ 0 & 0 \end{pmatrix} 
\,,\quad
H^{\mu\nu} = \begin{pmatrix} h^{ij} & 0 \\ 0 & 0 \end{pmatrix} 
\,,\quad
B_{\mu\nu} = \begin{pmatrix} 0 & -m_i \\ m_j & 0 \end{pmatrix} \,,
\label{NCKHB}
\ee
Note that the $U(1)$ gauge field $m_i$ appears in the off-diagonal components of the $B$-field, and therefore its $U(1)$ symmetry is induced in this picture by the gauge transformations $\delta B_{\mu v} = \partial_\mu \lambda_v$. 
Now, an obvious basis of null vectors would be $(x_\mu^I, y^\mu_I)$, with $I=1,2$ given by: 
\be
x_\mu^1 = \begin{pmatrix} \tau_i \\ 0 \end{pmatrix} \,,\quad
x_\mu^2 = \begin{pmatrix} 0 \\ 1 \end{pmatrix} \,,\quad
y^\mu_1 = \begin{pmatrix} -v^i \\ 0 \end{pmatrix} \,,\quad
y^\mu_2 = \begin{pmatrix} 0 \\ 1 \end{pmatrix} \,,\quad
\ee
such that
\be
Z_\mu{}^\nu =
\begin{pmatrix}
0 & \tau_i \\
- v^j & 0 
\end{pmatrix} \equiv 
 x_\mu^I \sigma_I{}^{I^\prime} y^\nu_{I^\prime} \,,\quad
\sigma_I{}^{I^\prime} = \begin{pmatrix} 0 & 1 \\ 1 & 0 \end{pmatrix} \,.
\ee
We can diagonalise this to match the canonical form of the generalised metric parametrisation of \cite{Morand:2017fnv}:
\be
Z_\mu{}^\nu = x_\mu y^\nu - \bar x_\mu \bar y^\nu \,,
\ee
with
\be
x_\mu = \frac{1}{\sqrt{2}} \begin{pmatrix} \tau_i \\ 1 \end{pmatrix} \,,\quad
\bx_\mu = \frac{1}{\sqrt{2}}\begin{pmatrix} \tau_i \\ -1 \end{pmatrix} \,,\quad
y^\mu = \frac{1}{\sqrt{2}} \begin{pmatrix} - v^i \\  1 \end{pmatrix} \,,\quad
\by^\mu = \frac{1}{\sqrt{2}} \begin{pmatrix} - v^i \\ - 1 \end{pmatrix} \,.
\label{NCxybasis}
\ee
We shall use this basis below.

\subsection{Bosonic worldsheet action}

\subsubsection*{Doubled action}

Now we will describe the doubled string action that we will use to obtain an action for a string in the Newton-Cartan background.
The starting point is the string worldsheet action in Hamiltonian form:\footnote{In this paper, as in \cite{Blair:2013noa}, the string tension is $T =1$.}
\be
 S = \int d^2\sigma \, \dot{X}^\mu P_\mu - \mathrm{Ham}(X,P) \,,
\ee
where
\be
\mathrm{Ham}(X,P) = \frac{e}{2}  Z^M \cH_{MN} Z^N + \frac{u}{2} Z^M \eta_{MN} Z^N \,,\quad Z^M \equiv \begin{pmatrix} X^{\prime \mu} \\ P_\mu \end{pmatrix} \,.
\ee
Here $e$ and $u$ are the two independent components of the worldsheet metric, imposing the vanishing of the string Hamiltonian constraints.
These constraints are written naturally in terms of doubled quantities.
One can define dual coordinates by $\tilde X_\mu^\prime = P_\mu$ (a prime denotes the worldsheet spatial derivative); integrating by parts we arrive at the doubled action of Tseytlin \cite{Tseytlin:1990nb, Tseytlin:1990va} written in terms of $X^M = ( X^\mu, \tilde X_\mu)$:\footnote{As discussed in \cite{Blair:2013noa}, there may be some subtleties related to the zero modes. We will ignore such subtleties here.}
\be
 S = \int d^2\sigma\, \frac{1}{2} \dot X^M \eta_{MN} X^{\prime N} - \frac{e}{2} X^{\prime M} \cH_{MN}(X) X^{\prime N}- \frac{u}{2} X^{\prime M} \eta_{MN} X^{\prime N} \,.
\label{bosonicdoubled}
\ee
In principle, we allow the generalised metric $\cH_{MN}(X)$ to depend on any of the $X^M$, however we impose the section constraint:
\be
\eta^{MN} \partial_M \cH_{PQ} \partial_N \cH_{KL} = 0 \,,
\ee
which guarantees closure of the algebra of \emph{worldsheet} diffeomorphisms \cite{Blair:2013noa}, and restricts us to backgrounds where we only depend on the usual number of coordinates.
One can view the choice of which half of the $X^M$ we allow the background to depend on as an expression of the manifest $O(D,D)$ covariance of this approach.
When the background has $N$ isometries, there is an ambiguity in the choice of which $X^M$ are chosen as the physical coordinates, and we obtain a true $O(N,N)$ T-duality symmetry.

\subsubsection*{String action in $(n,\bar n)$ background}

We can now consider the action \eqref{bosonicdoubled} for the background specified by the Newton-Cartan generalised metric - in fact, it is no more trouble to evaluate it on the general $(n,\bar n)$ parametrisation and then specify to Newton-Cartan at the end. 
We will find the result obtained by \cite{Morand:2017fnv} (who used an alternative but equivalent form of the doubled string action), but it is worth outlining the general procedure for completeness.
Readers solely interested in the immediate application to the Newton-Cartan geometry described above can simply mentally delete the indices $a$ and $\bar a$ everywhere they appear below.

First we run the doubling argument backwards: assuming the background only depends on $X^\mu$, we integrate the term in the doubled action \eqref{bosonicdoubled} involving $\dot X^M$ by parts so that it becomes $\dot{X}^\mu \tilde X^\prime_{\mu}$. Then $\tilde X$ only appears in the Lagrangian with a sigma derivative; we therefore let $P_\mu = \tilde X^\prime_\mu$ and will seek it to integrate this out of the action. 
(This total derivative could also be thought of as being cancelled against a ``topological'' term which can be added to the doubled worldsheet action as in \cite{Hull:2006va}. This term is not relevant classically, however it is important in the quantum doubled string, which we will not however consider in this paper.)

It is convenient to consider factoring out the $B$-field dependence:
\be
\mathcal{H} = U^T \uH U \,,\quad
\eta = U^T \eta U \,,
\quad
U \equiv \begin{pmatrix} 1 & 0 \\ - B & 1 \end{pmatrix} \,.
\label{factoroutB}
\ee
On all terms except $\dot{X}^\mu P_\mu$ this amounts to sending $P_\mu \rightarrow P_\mu - B_{\mu\nu} X^{\prime \nu}$.
Therefore redefining $\tilde P_\mu = P_\mu - B_{\mu\nu}X^{\prime \mu}$ we find the action is
\be
\begin{split}
	S & = \int d^2\sigma\, - \frac{e}{2} K_{\mu\nu} X^{\prime \mu}X^{\prime \nu} 
	+  B_{\mu\nu} \dot{X}^\mu X^{\prime \nu} 
	 - \frac{e}{2} H^{\mu\nu}  \tilde P_\mu \tilde P_\nu + \tilde P_\mu \mathcal{C}^\mu\,,
\end{split}
\label{LXinter}
\ee
where we used the fact that we have a $(n,\bar n)$ generalised metric with
\be
\uH_{MN}= \begin{pmatrix} K_{\mu\nu} & Z_\mu{}^\nu \\ Z_\nu{}^\mu & H^{\mu\nu} \end{pmatrix} \,,
\ee
and have
\be
\mathcal{C}^\mu = \dot{X}^\mu - u X^{\prime \mu} - e Z_\nu{}^\mu X^{\prime \nu} 
\label{C}
\,.
\ee
Note that we have generated the standard $B$-field coupling in \eqref{LXinter} by virtue of the redefinition from $P_\mu$ to $\tilde P_\mu$.

As $H^{\mu\nu}$ is not invertible, in order to integrate out $\tilde P_\mu$ we proceed as follows. The completeness relation implies
\be
\tilde P_\mu = K_{\mu\rho} H^{\rho \nu} \tilde P_\nu + x_\mu^a y^\nu_a \tilde P_\nu + \bar x_\mu^{\ba} \bar y^\nu_{\bar a} \tilde P_\nu 
\ee
so we let 
\be
A_\mu \equiv K_{\mu\rho} H^{\rho \nu} \tilde P_\nu \,\quad
\beta_{a} \equiv y^\nu_a \tilde P_\nu  \,\quad
\bar \beta_{\bar a} \equiv \bar y^{\nu}_{\bar a} \tilde P_\nu \,.
\ee
We insert this decomposition of $\tilde P_\mu$ into \eqref{LXinter} and add Lagrange multipliers to enforce the constraints $y^\mu_a A_\mu = 0 = \bar y^\mu_{\bar a} A_\mu$.
The part of the Lagrangian involving $\tilde P_\mu$ is then
\be
 - \frac{e}{2} H^{\mu\nu} A_\mu A_\nu + A_\mu ( \mathcal{C}^\mu - \Lambda_a y^\mu_a - \bar \Lambda_{\bar a} \bar y^\mu_{\bar a} ) + \beta_a x_\mu^a \mathcal{C}_\mu + \bar \beta_{\bar a} \bar x_\mu{}^{\bar a} \mathcal{C}_{\mu} \,.
\ee
The equation of motion for $A_\mu$ is 
\be
- e H^{\mu\nu} A_\nu + \mathcal{C}^\mu - \Lambda_a y^\mu_a - \bar \Lambda_{\bar a} \bar y^\mu_{\bar a} = 0 \,.
\label{eomA}
\ee
Contracting with $x_\mu^a$ and $\bar x_\mu^{\bar a}$ implies that $\Lambda_a = x_\mu^a \mathcal{C}^\mu$ and $\bar \Lambda_{\bar a} = \bar x_\mu^{\bar a} \mathcal{C}^\mu$.
Then contracting with $K_{\mu\nu}$ implies that
\be
A_\mu = \frac{1}{e} K_{\mu\nu} \mathcal{C}^\nu \,,
\ee
which indeed solves \eqref{eomA} given the solutions for the Lagrange multipliers.
We finally backsubstitute to find that \eqref{LXinter} becomes
\be
\begin{split} 
	S & = \int d^2\sigma\, \frac{1}{2e} K_{\mu\nu} \mathcal{C}^\mu \mathcal{C}^\nu 
	- \frac{e}{2} K_{\mu\nu} X^{\prime \mu} X^{\prime \nu} 
	+ B_{\mu\nu}  \dot X^{ \mu} X^{\prime \nu} 
	+  \beta_a x_\mu^a \mathcal{C}^\mu + \bar \beta_{\bar a} \bar x_\mu^{\bar a} \mathcal{C}^\mu 
	\\ & = \int d^2\sigma\,
	\frac{1}{2} K_{\mu\nu} \left( \frac{1}{e} ( \dot{X}^\mu - u X^{\prime \mu} ) ( \dot{X}^\nu - u X^{\prime \nu} )  - e X^{\prime \mu} X^{\prime \nu} \right) 
	+ B_{\mu\nu}  \dot X^{ \mu} X^{\prime \nu} 
	\\ & \qquad\qquad
	+  \beta_a x_\mu^a ( \dot{X}^\mu - (u+e) X^{\prime \mu})  + \bar \beta_{\bar a} \bar x_\mu^{\bar a} ( \dot{X}^\mu - ( u - e) X^{\prime \mu} ) \,.
\end{split}
\ee
This is written covariantly as:
\be
\begin{split} 
S & = \int d^2\sigma - \frac{1}{2} \sqrt{-\gamma} \gamma^{\alpha \beta} K_{\mu\nu} \partial_\alpha X^\mu \partial_\beta X^\nu - \frac{1}{2} \epsilon^{\alpha \beta} B_{\mu\nu} \partial_\alpha X^\mu \partial_\beta X^\nu \\
 & \qquad\qquad 
 + \beta_{\alpha a} x_\mu^a ( \sqrt{-\gamma} \gamma^{\alpha \beta} - \epsilon^{\alpha \beta} ) \partial_\beta X^\mu
 + \bar \beta_{\alpha \bar a} \bx_\mu^{\bar a} (- \sqrt{-\gamma} \gamma^{\alpha \beta} - \epsilon^{\alpha \beta} ) \partial_\beta X^\mu  \,.
 \label{covariantbosonic}
\end{split}
\ee
We have the standard kinetic term, except with the degenerate metric $K_{\mu\nu}$, as well as the standard $B$-field coupling, and have identified
\be
- \beta_{0a} + (u - e) \beta_{1a} \equiv \beta_a\,,\quad
- \bar \beta_{0\bar a} + (u + e) \bar \beta_{1\bar a} \equiv \bar\beta_{\bar a}\,.
\ee

\subsubsection*{Recovering the Newton-Cartan Polyakov action}

We take our coordinates to be $X^\mu = ( X^i, V)$ such that the duals are $\tilde X^\mu = ( \tilde X_i, U)$, where $U$ corresponds to the original null isometry direction of the metric \eqref{metric1}.
The background is specified by \eqref{NCKHB} and \eqref{NCxybasis}.
Inserting this into the action \eqref{covariantbosonic}, one finds:\footnote{Note the conventions $\epsilon^{01} = \epsilon_{01} = -1$, such that $\epsilon_{\alpha \gamma} \epsilon^{\beta \gamma} = \delta_\alpha^\beta$, $ (\det \gamma) \gamma^{\alpha \beta} \epsilon_{\beta \gamma} \gamma^{\gamma \delta} = \epsilon^{\alpha \beta}$.}
\be
\begin{split} 
S & = \int d^2\sigma - \frac{1}{2} \sqrt{-\gamma} \gamma^{\alpha \beta} h_{ij} \partial_\alpha X^i \partial_\beta X^j + \epsilon^{\alpha \beta} m_i \partial_\alpha X^i \partial_\beta V \\
 & \qquad\qquad 
 + \frac{1}{\sqrt{2}} \left( ( \beta_\alpha - \bar \beta_\alpha ) + \sqrt{-\gamma} \gamma^{\gamma \delta} \epsilon_{\gamma \alpha}(\beta_\delta+ \bar \beta_\delta) \right) ( \sqrt{-\gamma} \gamma^{\alpha \beta} \partial_\beta X^i \tau_i - \epsilon^{\alpha \beta} \partial_\beta V )
  \label{NCPolyakov}
\end{split}
\ee
This gives the Polyakov action for a string in Newton-Cartan gravity proposed in \cite{Harmark:2018cdl}. Note we can make things look simpler by defining $A_\alpha \equiv \frac{1}{\sqrt{2}}( ( \beta_\alpha - \bar \beta_\alpha ) + \sqrt{-\gamma} \gamma^{\gamma \delta} \epsilon_{\gamma \alpha}(\beta_\delta+ \bar \beta_\delta))$ to be the Lagrange multiplier. Solving the $V$ equation of motion by setting $A_\alpha -m_\alpha= -\partial_\alpha U$ returns us to the action for a string in the original Lorentzian background \eqref{metric1}. From the doubled perspective, we could have alternatively integrated out $V$ to obtain this.

\subsubsection*{Fradkin-Tseytlin term}

Let us add here a brief digression on the Fradkin-Tseytlin term describing the coupling of the generalised dilaton to the doubled worldsheet \cite{Hull:2006va}. 
\be
S_{FT} = \frac{1}{4\pi} \int d^2\sigma \sqrt{-\gamma}\, R^{(2)}(\gamma) \,d\,.
\ee
This is relevant when going beyond the classical action and treating the quantum doubled string. 
Adapting the standard arguments \cite{Buscher:1987sk, Hull:2006va}, integrating out the gauge fields $A_\mu$ in the path integral should generate a shift 
\be
d \rightarrow d - \frac{1}{4} \log \det{}^{\prime} ( H^{\mu\nu} )
\ee
where $\det{}^\prime$ denotes we should be taking this determinant on the restriction of the degenerate matrix $H^{\mu\nu}$ to the subspace spanned by the $A_\mu$ on which it is non-degenerate.
Now, for Newton-Cartan, in fact
\be
d = \frac{1}{4} \log \left|  \frac{1}{(d-1)!} \epsilon_{i_1 \dots i_d} \epsilon_{j_1 \dots j_d} v^{i_1} v^{j_1} h^{i_2 j_2} \dots h^{i_d j_d} \right|
\ee
is exactly $\frac{1}{4} \log \det{}^{\prime} ( H^{\mu\nu} )$, so that we get $d \rightarrow 0$ (in the absence of an original dilaton). This is consistent with what happens in a conventional background.

For the remainder of this paper, we consider only the classical worldsheet action, and its supersymmetrisation, and will not further discuss this Fradkin-Tseytlin term.
(We note that it would be relevant when considering the beta functional equations for the background fields, for instance, and we mention it both for completeness and with a view to further analysis of the quantisation of the Newton-Cartan string in the future.)

\subsection{Adding a $B$-field}
\label{addb}

As a slight extension of our previous procedure, let's outline what would happen if we assumed that our background \eqref{metric1} with a null isometry was also equipped with a B-field $B_{\mu\nu}$ with non-zero components $B_{ij}\equiv \mathsf{B}_{ij}$ and $ B_{i u} \equiv {B}_{i}$.
We consider the factorisation \eqref{factoroutB} for the original generalised metric. After T-dualising, the generalised metric admits the factorisation
\be
\cH = \tilde U \cH_{\text{NC}} \tilde U^T \,.
\ee
where $\cH_{\text{NC}}$ is the (original) Newton-Cartan generalised metric \eqref{NCGM}, for which we used the parametrisation in terms of $H,K,B$ and zero vectors given by \eqref{NCKHB} and \eqref{NCxybasis}, and $\tilde U$ can itself be factorised as
\be
\tilde U = U_{\mathsf{B}} U_A = U_A U_{\mathsf{B}} 
\,,\quad
(U_{\mathsf{B}})^M{}_N = \begin{pmatrix} 
\delta^\mu{}_\nu & 0 \\ 
- \mathsf{B}_{\mu\nu} & \delta_\mu{}^\nu 
\end{pmatrix}\,,\quad
(U_A)^M{}_N\equiv \begin{pmatrix}
A^\mu{}_\nu & 0 \\ 0 & (A^{-1})^{\nu}{}_\mu
\end{pmatrix} \,,
\ee
where the only non-zero components of $\mathsf{B}_{\mu\nu}$ are $\mathsf{B}_{ij}$,
and 
\be
A^\mu{}_\nu \equiv \begin{pmatrix} \delta^i{}_j & 0 \\ B_j & 1 \end{pmatrix} \,,
\ee
generates a geometric $\mathrm{GL}(d+1)$ transformation. Note that $A^\mu{}_\nu \mathsf{B}_{\mu \rho} = \mathsf{B}_{\nu \rho}$, $\mathsf{B}_{\mu\rho} A^{\rho}{}_\nu  = B_{\mu \nu}$.
Now, overall conjugation of $\cH_{\text{NC}}$ by $U_{\mathsf{B}}$ simply has the effect of turning on the components $\mathsf{B}_{ij}$, and does not materially change any aspect of our analysis.
The conjugation by $U_A$ on the other hand does have an impact, but this is easily calculated.
The overall result is that the parametrisation given by \eqref{NCKHB} and \eqref{NCxybasis} is replaced by the following one:
\be
K_{\mu\nu} = \begin{pmatrix} h_{ij} & 0 \\ 0 & 0 \end{pmatrix}  ,\quad
H^{\mu\nu} = \begin{pmatrix} h^{ij} & - h^{ik} B_k \\ -h^{jk} B_k & h^{kl} B_k B_l\end{pmatrix},\quad
B_{\mu\nu} = \begin{pmatrix}
\mathsf{B}_{ij} -2 m_{[i}B_{j]} & - m_i \\ m_j & 0 
\end{pmatrix} ,
\label{NCKHB_b}
\ee
with the zero vectors:
\be
x_\mu = 
\frac{1}{\sqrt{2}} \begin{pmatrix} 
\tau_i + B_i \\ 1 
\end{pmatrix} ,\quad
\bar x_\mu =
\frac{1}{\sqrt{2}} \begin{pmatrix}
\tau_i - B_i \\ - 1 
\end{pmatrix} ,\quad
y^\mu = \frac{1}{\sqrt{2}}
\begin{pmatrix}
-v^i \\ 1 + B_k v^k 
\end{pmatrix} ,\quad
\bar y^\mu = \frac{1}{\sqrt{2}}
\begin{pmatrix}
-v^i \\ - 1 + B_k v^k 
\end{pmatrix} .
\label{NCxybasis_b}
\ee
With this parametrisation, we see immediately that the Newton-Cartan action \eqref{NCPolyakov} becomes that of \cite{Harmark:2019upf}:
\be
\begin{split} 
S & = \int d^2\sigma - \frac{1}{2} \sqrt{-\gamma} \gamma^{\alpha \beta} h_{ij} \partial_\alpha X^i \partial_\beta X^j
- \frac{1}{2} \epsilon^{\alpha \beta} \mathsf{B}_{ij} \partial_\alpha X^i \partial_\beta X^j
 + \epsilon^{\alpha \beta} m_i \partial_\alpha X^i (\partial_\beta V + B_i \partial_\beta X^i) \\
 & \qquad\qquad 
 + \frac{1}{\sqrt{2}} \left( ( \beta_\alpha - \bar \beta_\alpha ) + \sqrt{-\gamma} \gamma^{\gamma \delta} \epsilon_{\gamma \alpha}(\beta_\delta+ \bar \beta_\delta) \right) \left( \sqrt{-\gamma} \gamma^{\alpha \beta} \partial_\beta X^i \tau_i - \epsilon^{\alpha \beta} ( \partial_\beta V + B_i \partial_\beta X^i) \right)\,.
  \label{NCPolyakovB}
\end{split}
\ee

\section{Vielbeins, worldsheet fermions and duality}
\label{vielbeininterlude} 

The goal of this section is to introduce some necessary technology in the form of doubled pseudo-vielbeins for the projectors \eqref{projs}, and to discuss in general terms some features of the description of worldsheet fermions in the Newton-Cartan background.

\subsection{Doubled vielbeins} 
\label{vieltech} 

In the standard RNS string, the worldsheet bosons $X^\mu$ are accompanied by their worldsheet superpartners in the form of a pair of Majorana-Weyl fermions of opposite chirality, $\psi^\mu$ and $\tilde\psi^\mu$. How should we describe these fermions in an $O(D,D)$ covariant picture? 

For the bosons, the idea was to first go to the Hamiltonian setting, pairing the coordinates $X^\mu$ with their momenta $P_\mu$. This provided a natural doubling.
On the other hand, fermions are already their own momenta (their kinetic term will be $\sim \psi \dot{\psi}$) and so they should not be doubled in the same way.
This suggests exchanging $(X^\mu, P_\mu) \rightarrow X^M$ but continuing to work with the original fermions $\psi^\mu$ and $\tilde\psi^\mu$. The latter are spacetime vectors. Generically, there is no canonical way to express a single spacetime vector in an $O(D,D)$ covariant manner. However, as the fermions come with different chirality and so are naturally left- and right-moving, there is a natural way to associate them to the denominator subgroup in the coset $O(D,D)/O(1,D-1) \times O(1,D-1)$, which consists of copies of the Lorentz group seen separately by left- and right-movers on the worldsheet. This motivates defining the fermions with flat indices $\psi^{\mA}$ and $\tilde\psi^{\pA}$ such that $\psi^\mu = e^\mu{}_{\mA} \psi^{\mA}$, $\tilde\psi^\mu = e^\mu{}_{\pA} \tilde\psi^{\pA}$, with $\psi^{\mA}$ and $\tilde\psi^{\pA}$ transforming under the separate Lorentz group factors (with associated different flat indices $\mA$ and $\pA$ for emphasis, and in principle we could use separate vielbeins in each sector, as will appear naturally after T-dualising). Crucially though, these fermions with flat indices do not transform under the global $O(D,D)$.

In fact, we could equivalently define doubled fermions $\psi^M$ and $\tilde \psi^M$ as follows. (This is what is used in \cite{HackettJones:2006bp, Hull:2006va} in an alternative approach to the worldsheet supersymmetric doubled string.)
The definition of the generalised metric in \eqref{gmconds} implies the existence of projectors:
\be
P^M{}_N = \frac{1}{2} ( \delta^M{}_N  + \eta^{MK} \mathcal{H}_{KN} ) \,,\quad
\bar P^M{}_N = \frac{1}{2} ( \delta^M{}_N  - \eta^{MK} \mathcal{H}_{KN} ) \,.
\label{projs}
\ee
which can be thought of as projecting onto the separate $D$ dimensional subspaces associated to each doubled Lorentz factor. Requiring
\be
\Pp^M{}_N \psi^N = 0 \,,\quad \Pm^M{}_N \psi^N = \psi^M \,,\quad
\Pm^M{}_N \tilde\psi^N = 0 \,,\quad \Pp^M{}_N \tilde \psi^N = \tilde \psi^M \,,\quad
\label{ppsireq}
\ee
implies that each of $\psi^M$ and $\tilde\psi^M$ only have $D$ independent components (for $n=\bar n$). 
In the standard parametrisation, for instance, this gives
\be
\psi^M = \frac{1}{\sqrt{2}} \begin{pmatrix}
\psi^\mu \\ (-g+B)_{\mu\nu} \psi^\nu 
\end{pmatrix} \,,\quad
\tilde\psi^M = \frac{1}{\sqrt{2}} \begin{pmatrix}
\tilde \psi^\mu \\ (g+B)_{\mu\nu} \tilde\psi^\nu 
\end{pmatrix} \,.
\ee 
We can connect these two pictures by solving the conditions \eqref{ppsireq} by writing $\psi^M = \Vm^M{}_{\mA} \psi^{\mA}$ and $\tilde \psi^M = \Vp^M{}_{\pA} \psi^{\pA}$, where $\Vm^M{}_{\mA}$ and $\Vp^M{}_{\pA}$ can be constructed as ``vielbeins'' for the projectors themselves (we will for convenience refer to all these non-square pseudo-vielbeins simply as vielbeins, to avoid awkward phrasing).
To define these in full generality, let's assume again we have a general $(n,\bar n)$ parametrisation of the generalised metric.
This actually means \cite{Morand:2017fnv} that the full doubled Lorentz group (i.e. the denominator subgroup in the coset that the generalised metric parametrises) is $O(t+n,s+n) \times O(t+\bar n, s+\bar n)$, with $t+s+n+\bar n = D$. 
So now $\mA = 1, \dots, D+\bar n - n$ and $\pA = 1, \dots, D+n-\bar n$ are the corresponding flat indices.

Then we introduce $\Vm^M{}_{\mA}$ and $\Vp^M{}_{\pA}$ such that (note the sign in the first expression):
\be
\begin{split} 
	\Pm_{MN} & = \frac{1}{2} ( \eta_{MN} - \cH_{MN} )= -\Vm_M{}^{\mA} \Vm_N{}^{\mB} \flathm_{\mA \mB} \\
	\Pp_{MN} & = \frac{1}{2} ( \eta_{MN} + \cH_{MN}  )= \Vp_M{}^{\pA} \Vp_N{}^{\pB} \flathp_{\pA \pB} \,,
\end{split}
\label{projviel}
\ee
where $\flathm_{\mA \mB}$ and $\flathp_{\pA \pB}$ are $O(t+\bar n, s+\bar n)$ and $O(t+n,s+n)$ flat metrics, respectively. These obey various identities:
\be
\Pp_M{}^N \Vm_N{}^{\mA} = 0 \,,\quad \Pm_M{}^N \Vm_N{}^{\mA} = \Vm_M{}^{\mA} \,,\quad 
\Pm_M{}^N \Vp_N{}^{\pA} = 0 \,,\quad \Pp_M{}^N \Vp_N{}^{\pA} = \Vp_M{}^{\pA} \,, 
\ee
\be
\eta^{MN} \Vm_M{}^{\mA} \Vm_N{}^{\mB} = - \flathm^{\mA \mB} \,,\quad
\eta^{MN} \Vp_M{}^{\pA} \Vp_N{}^{\pB} = + \flathp^{\pA \pB} \,,\quad
\eta^{MN} \Vm_M{}^{\mA} \Vp_N{}^{\pB} = 0 \,.
\ee
The paper \cite{Morand:2017fnv} introduced the following explicit parametrisation.
Decompose the flat indices as $\mA = ( \bar m, \bar a, \bar a)$ and $\pA = ( m, a,a)$, where $m$ and $\bar m$ are $D-n-\bar n$ dimensional flat indices, and $a$ and $\bar a$ are the indices corresponding to the zero vectors appearing in the generalised metric. 
Let's pre-emptively point out that for Newton-Cartan, one can ignore the indices $a$ and $\bar a$ (as they are one-dimensional), and we will in fact not distinguish between the $D-2$ dimensional flat indices $m$ and $\bar m$, but denote both instead by\footnote{The author apologises for this, and also for the sheer number of versions of ``h'' in use. The flat matrices $\boldsymbol{h}_{AB}$, $\flathm_{\mA \mB}$ are not to be confused with the degenerate $h_{ij}$, $h^{ij}$ appearing in the Newton-Cartan geometry (nor, for that matter, are they to be confused with anything else).} $\um$.
Introduce non-square ``vielbeins'' for the degenerate matrices $K$ and $H$ involving flat $O(t,s)$ metrics $\eta_{mn}$ and $\bar\eta_{\bar m \bar n}$:
\be
K_{\mu\nu} = k_\mu{}^m k_\nu{}^n \eta_{mn} = \bar k_{\mu}{}^{\bar m} \bar k_{\nu}{}^{\bar n} \eta_{\bar m \bar n} \,,\quad
H^{\mu\nu} = h^{\mu}{}_m h^\nu{}_n \eta^{mn} = \bar h^\mu{}_{\bar m} \bar h^{\nu}{}_{\bar n} \eta^{\bar m \bar n} \,,
\ee
where $k_\mu{}^m y^\mu_a = \bar k_\mu{}^{\bar m} \by^\mu_{\bar a} = 0$, $h^\mu{}_m x_\mu^a = \bar h^\mu{}_{\bar m} \bx_\mu^{\bar a} = 0$, and we have completeness relations
\be
k_\mu{}^m h^\mu{}_n = \delta^m_n\,,\quad
\bar k_\mu{}^{\bar m} \bar h^{\mu}{}_{\bar n} = \delta^{\bar m}_{\bar n} \,,
\ee
\be
k_\mu{}^m h^\nu{}_m + x_\mu^a y^\nu_b + \bx_\mu^{\ba} \by^\nu_{\bb} = \delta_\mu^\nu \,,\quad
\bar k_\mu{}^{\bar m} \bar h^\nu{}_{\bar n} + x_\mu^a y^\nu_b + \bx_\mu^{\ba} \by^\nu_{\bb} = \delta_\mu^\nu \,.
\ee
Define
\be
k_\mu{}^{\pA} \equiv \begin{pmatrix} k_\mu{}^m & x_\mu^a & x_\mu^a \end{pmatrix} \,,\quad
\bar k_\mu{}^{\mA} \equiv \begin{pmatrix} \bar k_\mu{}^{\bar m} & \bx_\mu^{\bar a} & \bx_\mu^{\bar a} \end{pmatrix} \,,
\ee
\be
h^\mu{}_{\pA} \equiv \begin{pmatrix} h^{\mu}{}_m & y^\mu_a & y^\mu_a \end{pmatrix} \,,\quad
\bar h^\mu{}_{\mA} \equiv \begin{pmatrix} \bar  h^\mu{}_{\bar m}  & \by^\mu_{\bar a} & \by^\mu_{\bar a} \end{pmatrix} \,,
\ee
and then let 
\be
\Vm_{M \mA} = \frac{1}{\sqrt{2}} \begin{pmatrix} - \bar k_{\mu \mA} + B_{\mu\nu} \bar h^{\nu}{}_{\mA} \\ \bar h^{\mu}{}_{\mA} \end{pmatrix} 
\,,\quad
\Vp_{M \pA} = \frac{1}{\sqrt{2}} \begin{pmatrix}  k_{\mu \pA} + B_{\mu\nu}  h^{\nu}{}_{\pA} \\  h^{\mu}{}_{\pA} \end{pmatrix} \,.
\ee
The flat indices are raised and lowered using 
\be
\flathm_{\mA \mB} = \begin{pmatrix}
\eta_{\bar m \bar n} & 0 & 0 \\ 0 & - \delta_{\bar a \bar b} & 0 \\ 0 & 0 & \delta_{\bar a \bar b} 
\end{pmatrix} 
\,,\quad
\flathp_{\pA \pB} = \begin{pmatrix}
\eta_{ m  n} & 0 & 0 \\ 0 & - \delta_{ a  b} & 0 \\ 0 & 0 & \delta_{ a  b} 
\end{pmatrix} \,.
\ee
In fact, in this paper we will work with a perhaps slightly simpler parametrisation where we off-diagonalise the blocks in these flat metrics, leading to:
\be
\flathm_{\mA \mB} = \begin{pmatrix}
\eta_{\bar m \bar n} & 0 & 0 \\ 0 & 0 &  \delta_{\bar a \bar b}\\ 0 &  \delta_{\bar a \bar b} & 0
\end{pmatrix} 
\,,\quad
\flathp_{\pA \pB} = \begin{pmatrix}
\eta_{ m  n} & 0 & 0 \\ 0 & 0 &  \delta_{ a  b} \\ 0 & \delta_{ a  b}  & 0
\end{pmatrix} \,,
\ee
such that
\be
\bar k_\mu{}^{\mA} \equiv \begin{pmatrix} \bar k_\mu{}^{\bar m} & \sqrt{2} \bx_\mu^{\bar a} &0 \end{pmatrix} 
\,,\quad
k_\mu{}^{\pA} \equiv \begin{pmatrix} k_\mu{}^m &  \sqrt{2} x_\mu^a & 0 \end{pmatrix} 
\,,
\ee
\be
\bar h^\mu{}_{\mA} \equiv \begin{pmatrix} \bar  h^\mu{}_{\bar m}  &\sqrt{2} \by^\mu_{\bar a} & 0\end{pmatrix} 
\,,\quad
h^\mu{}_{\pA} \equiv \begin{pmatrix} h^{\mu}{}_m & \sqrt{2} y^\mu_a & 0 \end{pmatrix} 
\,.
\ee
It is useful to record that in both cases we have
\be
H^{\mu\nu} k_\nu{}^{\pA} + x_\nu^b y^\mu_b h^{\nu \pA} = h^{\mu \pA} \,,\quad
K_{\mu\nu} h^{\nu}{}_{\pA} + x_\mu^b y^\nu_b k_{\nu \pA} = k_{\mu \pA} \,,
\label{usefulhk1}
\ee
\be
H^{\mu\nu} \bar k_\nu{}^{\mA} + \bx_\nu^{\bb} \by^\mu_{\bb} \bar h^{\nu \mA} = \bar h^{\mu \mA} \,,\quad
K_{\mu\nu} \bar h^{\nu}{}_{\mA} + \bx_\mu^{\bb} \by^\nu_{\bb} \bar k_{\nu \mA} = \bar k_{\mu \mA} \,.
\label{usefulhk2}
\ee

\subsection{Newton-Cartan vielbeins}
\label{NCvieltech}

Let's now discuss in general terms what we expect to happen when we consider the supersymmetric Newton-Cartan string, based on expectations from T-duality on the worldsheet. 

\subsubsection*{T-duality on spacetime vielbeins}

First of all, let's suppose we are dealing with an ordinary supergravity background, for which we introduce a vielbein $e^A{}_\mu$ such that $g_{\mu\nu} = e^A{}_\mu e^B{}_\nu \flathp_{AB}$.
Then the projector vielbeins are (identifying the vielbeins and hence flat indices in each):
\be
V_{M A} = \frac{1}{\sqrt{2}} \begin{pmatrix} e_{\mu A} + B_{\mu\nu} e^\nu{}_A \\ e^\mu{}_A \end{pmatrix} \,,\quad
\bar V_{M A} = \frac{1}{\sqrt{2}} \begin{pmatrix} - e_{\mu A} + B_{\mu\nu} e^\nu{}_A \\ e^\mu{}_A \end{pmatrix} \,.
\ee
Let's split $\mu = (i,z)$ and carry out a Buscher transformation. One finds \cite{Hassan:1999bv} that the spacetime vielbein $e^A{}_\mu$ transforms differently in the left and right projected sectors.
There are two transformations (here the $+$ case is associated to $\Vp$ and the $-$ case to $\Vm$):
\be
e_\pm^\mu{}_A = Q_\pm^\mu{}_\nu e^\nu{}_A \,,\quad 
Q_\pm^\mu{}_\nu = \begin{pmatrix} \delta^i{}_j & 0 \\ \pm g_{zj} + B_{zj} & \pm g_{zz} \end{pmatrix} \,,
\ee
and the different vielbein are related by a Lorentz transformation:\footnote{For a conventional choice of vielbein, with $A=(\underline i, \underline z)$, we have $e_z{}^{\underline i} = 0$, $e_z{}^{\underline z} = \sqrt{g_{zz}}$, then $\Lambda^{\underline i}{}_{\underline j} = \delta^{\underline i}{}_{\underline j}$, $\Lambda^{\underline z}{}_{\underline z} = -1$, and this gives the standard T-duality minus sign flip, $\psi^{\underline{i}} \rightarrow \psi^{\underline{i}}$, $\psi^{\underline z} \rightarrow - \psi^{\underline z}$, for example.}
\be
e_+^\mu{}_A = e_-^\mu{}_B \Lambda^B{}_A \,,\quad \Lambda^B{}_A \equiv e^B{}_\nu (Q_-^{-1})^\nu{}_\rho  Q_+^\rho{}_\mu e^\mu{}_A = \delta^B_A - 2 \frac{1}{g_{zz}} e_z{}^B e_{z A} \,.
\ee
Evidently, if $g_{zz} = 0$, the transformations $Q_\pm$ relating $e$ to $e_\pm$ are non-invertible, and the different vielbein can not be related to each other.

\subsubsection*{Null duality on Newton-Cartan vielbein} 

For the Lorentzian metric \eqref{metric1}, a natural choice of vielbein is to write\footnote{We adopt the convention of writing flat index components in the Newton-Cartan background with an underline.}
\be
ds^2 = 2 e^{\utau} e^{\uu} + \delta_{\um\un} e^{\um}{}_i e^{\un}{}_j dx^i dx^j\,,\quad e^{\utau} \equiv \tau_i dx^i \,,\quad e^{\uu} \equiv du - m_i dx^i\,,
\ee
where we introduce a non-square ``vielbein'' $e^{\um}{}_i$ such that
\be
h_{ij} = \delta_{\um\un} e^{\um}{}_i e^{\un}{}_j
\ee
such that $(e_i{}^{\um}, \tau_i)$ is invertible and has inverse $(e^i{}_{\um}, - v^i)$ \cite{Hansen:2018ofj}.
We have identities 
\be
e_i{}^{\um} e_{\um}{}^j =\delta_i^j + \tau_i v^j \,,\quad e_{\um}{}^j e_j{}^{\un} = \delta_{\um}{}^{\un}\,,
\ee
with $h^{ij} = e^i{}_{\um} e^j{}_{\un} \delta^{\um\un}$. 
The full vielbein is then
\be
e^A{}_\mu = \begin{pmatrix} e^{\um}{}_i & 0 \\ \tau_i & 0 \\ -m_i & 1 \end{pmatrix} \,,\quad
e^\mu{}_A = \begin{pmatrix} e_{\um}{}^i & - v^i & 0 \\ e_{\um}{}^j m_j & - v^j m_j & 1 \end{pmatrix} 
\,,\quad
\flathp_{AB} = \begin{pmatrix} \delta_{\um\un} & 0 & 0 \\ 0  & 0 & 1 \\ 0 & 1 & 0 \end{pmatrix} \,.
\ee
From this, the T-dual inverse vielbeins are:
\be
e_\pm^\mu{}_A = \begin{pmatrix} e^i{}_{\um} & - v^i & 0 \\ 0 & \pm 1 & 0 \end{pmatrix} 
\ee
which are as expected not invertible.
We can explicitly check that there is no $\Lambda^B{}_A$ such that $e_+^\mu{}_A = e_-^\mu{}_B \Lambda^B{}_A$, as if there were this would require $e^i{}_{\um} \Lambda^{\um}{}_{\utau} + v^i = - v^i$, and contracting with $\tau_i$ gives $-1 = 1$.
Nonetheless, the doubled vielbeins are well-defined: 
\be
V_{M \pA} = \frac{1}{\sqrt{2}} \begin{pmatrix} e_{\um i} & - m_i & \tau_i \\ e_{\um}{}^k m_k & - v^j m_j & 1 \\e^i{}_{\um} & - v^i & 0 \\ 0 &  1 & 0
\end{pmatrix} \,,\quad
\bar V_{M \mA} = \frac{1}{\sqrt{2}} \begin{pmatrix} -e_{\um i} &  m_i & -\tau_i \\ e_{\um}{}^k m_k & - v^j m_j & 1 \\e^i{}_{\um} & - v^i & 0 \\ 0 &  -1 & 0
\end{pmatrix} \,.
\ee
This is consistent with the parametrisation of section \ref{vieltech}, with
\be
k_\mu{}^{\pA} = \begin{pmatrix}  k_\mu{}^{\um} & \sqrt{2} x_\mu & 0 \end{pmatrix} \,,\quad
\bar k_\mu{}^{\mA} = \begin{pmatrix} k_\mu{}^{\um} & \sqrt{2} \bar x_\mu & 0 \end{pmatrix} \,,\quad
k_\mu{}^{\um} \equiv \begin{pmatrix} e_i{}^{\um} & 0 \end{pmatrix} \,,
\label{NCviel1}
\ee
\be
h^\mu{}_{\pA} = \begin{pmatrix}  h^\mu{}_{\um} & \sqrt{2} y^\mu & 0 \end{pmatrix} \,,\quad
\bar h^\mu{}_{\mA} = \begin{pmatrix} h^\mu{}_{\um} & \sqrt{2} \bar y^\mu & 0 \end{pmatrix} \,,\quad
h^\mu{}_{\um} \equiv \begin{pmatrix} e^i{}_{\um} & 0 \end{pmatrix} \,,
\label{NCviel2}
\ee
The way the relationship between $e_+$ and $e_-$ breaks down is clearly consistent with the fact that we no longer have a relativistic background geometry. 
We also remember that in the bosonic sector, we had what were essentially chirality conditions enforcing that the directions $x_\mu X^\mu$ and $\bar x_\mu X^\mu$ be either left- or right-moving only. 
The lack of a Lorentz transformation which can be used to align the left- and right-moving sectors of the worldsheet is then seen to be connected to these directions becoming chiral.
We see however from the above parametrisation that the left and right doubled vielbeins contain still the same $d \times (d-1)$ not-square ``vielbein'' $e_{\um}{}^i$, which means that in the $(d-1)$-dimensional subspace on which $h_{ij}$ is non-degenerate the string left- and right-moving sectors see the same space - in this case, they agree on the spatial hypersurfaces orthogonal to the Newton-Cartan time direction specified by $v^i$. Note that it is for this reason that we use the common flat index $\um$ for both $\Vm_{M\mA}$ and $\Vp_{M \pA}$, rather than separate indices $m$ and $\bar m$ as indicated in the previous subsection.

One might wonder as a result about how one should think of the separate chiral $O(1,d)$ groups under which $\psi^{\mA}$ and $\tilde \psi^{\pA}$ are said to transform. 
We can decompose $O(1,d)$ into $O(d-1)$ acting on the $\um$ indices, $O(1,1)$ transformations acting in the $(\utau,\uu)$ directions, with $2(d-1)$ ``mixed'' transformations leftover.
With the vielbein choices as above (viewed as a Lorentz gauge fixing), half of the latter survive and implement (finite) Galilean transformations via the action of
\be
\Lambda^A{}_{B} = \begin{pmatrix}
\delta^{\um}_{\un} & \lambda^{\um} & 0 \\ 0 & 1 & 0 \\ - \lambda_{\un} & -\frac{1}{2} \lambda^{\up} \lambda_{\up} & 1 
\end{pmatrix} \,,
\ee
which induces $e_i{}^{\um} \rightarrow e_i{}^{\um}+ \lambda^{\um} \tau_i$, $m_i \rightarrow m_i+ \lambda^{\um} e_{i \um} - \frac{1}{2}\lambda^{\up} \lambda_{\up} \tau_i$, $v^i \rightarrow v^i + e^{i}{}_{\um} \lambda^{\um}$. In the non-Riemannian parametrisations of the generalised metric \cite{Morand:2017fnv}, this transformation can be viewed as a shift symmetry acting on the decomposition into a particular $K_{\mu\nu}, y^\mu,\bar y^\mu$ and $B_{\mu\nu}$.

\subsubsection*{Worldsheet fermions} 

The RNS string action contains fermions $\psi^\mu, \tilde \psi^\mu$ with kinetic terms like
\be
L_\psi \sim \frac{i}{2} \psi^\mu g_{\mu\nu} \partial \psi^\nu \,.
\ee
T-dualising only explicitly changes the part of the action coupling directly to the bosonic coordinates $X^\mu$, $L_\psi$ itself is \emph{invariant} under T-duality in the trivial sense that one can find a change of variables $\psi^\mu \rightarrow \psi^\mu_{\text{dual}}$ such that
\be
\frac{i}{2} \psi^\mu g_{\mu\nu} \partial \psi^\nu = \frac{i}{2} \psi^\mu_{\text{dual}} (g_{\text{dual}})_{\mu\nu} \partial \psi_{\text{dual}}^\nu
\ee
with the metric transforming according to the Buscher rules.
As we discussed, a way to make this manifest is to flatten the indices on the fermions, such that $\psi^A = e^A{}_\mu \psi^\mu$, and declare that $\psi^A$ is an invariant under $O(D,D)$.
Note that in this case, 
\be
\psi^\mu_{\text{dual}} = e_\pm^\mu{}_A \psi^A = Q_{\pm}^\mu{}_\nu  e^\nu{}_A \psi^A = Q_\pm^\mu{}_\nu \psi^\nu \,.
\ee
Hence the worldsheet fermions with curved spacetime indices transform like the spacetime vielbein.

This picture breaks down when doing a null duality. In particular there is no well-defined notion of $\psi^\mu_{\text{dual}}$. This is because there is no well-defined (invertible) spacetime vielbein $e_\pm^\mu{}_A$ which can be extracted from the doubled vielbeins.
We can see this quite clearly by defining (for example) $\psi^M = V^M{}_A \psi^A$. We have 
\be
\psi^M = \frac{1}{\sqrt{2}} \begin{pmatrix} \psi^\mu \\ (g+B)_{\mu\nu} \psi^\nu \end{pmatrix} 
\,,\quad
\psi^M_{\text{dual}} = 
\frac{1}{\sqrt{2}} \begin{pmatrix} 
\psi^i \\ g_{zz} \psi^z + (g+B)_{zj} \psi^j \\
(g+B)_{ij} \psi^j + (g+B)_{iz} \psi^z \\
\psi^z 
\end{pmatrix}\,,
\ee
which would ordinarily lead to the expected results 
\be
\psi_{\text{dual}}^i = \psi^i \,,\quad \psi_{\text{dual}}^z = g_{zz} \psi^z + ( g+B)_{zj} \psi^j\,.
\ee
When $g_{zz} = 0$, $\psi^z$ drops out of the putative definition for $\psi_{\text{dual}}^\mu$ entirely. Nevertheless, it is of course still present in the doubled variable $\psi^M_{\text{dual}}$ or in the invariant $\psi^A$. This is just saying (again) that in such circumstances one cannot use the standard spacetime parametrisations and intuition, however starting with the doubled picture and then adopting the correct Newton-Cartan parametrisation will lead to sensible results.

Let's now consider working in the Lorentzian background with null isometry, prior to passing to the Newton-Cartan description.
Here we have a well-defined spacetime vielbein so it is equivalent to work with either worldsheet fermions with curved spacetime indices, which we denote $\psi^\mu$, or flat indices, $\psi^A = e^A{}_\mu \psi^\mu$.
Explicitly, we have
\be
\psi^A \equiv \begin{pmatrix}
\psi^{\underline m} \\ \psi^{\underline \tau} \\ \psi^{\underline{u}} 
\end{pmatrix}
= 
\begin{pmatrix}
e^{\underline m}{}_i \psi^i \\ \tau_i \psi^i \\ \psi^u - m_i \psi^i
\end{pmatrix} \,.
\label{psiApsimu}
\ee
Note for instance that:
\be
\begin{split}
g_{\mu\nu} \psi^\mu \partial \psi^\nu 
& = \bar h_{ij} \psi^i \partial \psi^j + \tau_j \psi^u \partial \psi^j + \tau_j \psi^j \partial \psi^u 
\\ & = \delta_{\underline m \underline n} \psi^{\underline m} \partial \psi^{\underline n} 
 + \psi^{\underline \tau} \partial \psi^{\uu} + \psi^{\underline u} \partial \psi^{\underline \tau} + \dots
\\ & = \flathp_{AB} \psi^A \partial \psi^B  + \dots\,,
\end{split}
\ee
where the ellipsis denotes derivatives of the background, which combine into terms featuring the spin connection of the background (see appendix \ref{ws}).
In going to the Newton-Cartan description using the null duality, we will write the action for the fermions in terms of the flat indexed quantities.

\section{Worldsheet supersymmetric Newton-Cartan string}
\label{NCRNS}

\subsection{The worldsheet supersymmetric doubled action}
\label{RNSaction} 

\subsubsection*{Details of the action}

The doubled RNS string of \cite{Blair:2013noa} extends the bosonic action \eqref{bosonicdoubled} to the worldsheet supersymmetric action:\footnote{We have slightly changed some of the (also surprising to the author) notation and conventions of \cite{Blair:2013noa}. In particular, our names for the doubled vielbeins are $\Vp \equiv R$ and $\Vm \equiv L$, and unlike in \cite{Blair:2013noa} we do not raise indices with the generalised metric but with the $O(D,D)$ structure.} 
\be
\begin{split} 
	S = \int d^2 \sigma \,& 
	 \frac{1}{2} \dot{X}^M \eta_{MN} X^{\prime N} - \frac{i}{2} ( \psi^{\mA} \dot{\psi}^{\mB} \flathm_{\mA \mB} + \tilde \psi^{\pA} \dot{\tilde{\psi}}^{\pB} \flathp_{\pA \pB} )
- \lambda \mathcal{H} 
- \tilde \lambda \tilde{\mathcal{H}}
- i\xi \mathcal{Q}
- i\tilde\xi \tilde{\mathcal{Q}}
\,.
\end{split} 
\label{bmr}
\ee
As we discussed above, the worldsheet fermions $\psi^{\mA}$ and $\tilde\psi^{\pA}$ carry \emph{flat} indices, associating each to one of the two separate factors of the doubled Lorentz group.\footnote{Note that this means that strictly speaking in order to show invariance of the action under spacetime Lorentz transformations, one has to make a certain non-local transformation of the $X^M$ which only affects the dual coordinates $\tilde X_\mu$, being of the form $X^M \rightarrow X^M + \int^\sigma d\sigma^\prime \partial^M \Lambda (\dots)$. As ultimately only $\tilde X^\prime_{\mu}$ appears in the action, this is sensible.} We now have both bosonic Lagrange multipliers, $\lambda$ and $\tilde \lambda$, related to the parametrisation of the worldsheet metric (with $\lambda = e-u$, $\tilde \lambda = e+u$), and fermionic Lagrange multipliers, $\xi$ and $\tilde \xi$, related to the parametrisation of the worldsheet gravitino, as detailed in appendix \ref{ws}.
They enforce the super-Virasoro constraints, $\cH = \tilde \cH = \mathcal{Q} = \tilde{\mathcal{Q}} = 0$.
To describe these, we first need to use the projector vielbeins of \eqref{projviel} to build objects which resemble spin connections,
\be
\omega_{M \mA \mB} = - \Vm_{N \mA} \partial_M \Vm^N{}_{\mB} + \Vm^N{}_{[\mA} \Vm^P{}_{\mB]} \partial_P \cH_{MN} \,,\quad
\tilde\omega_{M \pA \pB} = \Vp_{N \pA} \partial_M \Vp^N{}_{\pB} + \Vp^N{}_{[\pA} \Vp^P{}_{\pB]} \partial_P \cH_{MN} \,.
\ee
Though we will casually refer to these as spin connections, under generalised diffeomorphisms (the $O(D,D)$ covariantisation of spacetime diffeomorphisms and $B$-field gauge transformations) they do not transform as a (generalised) connection should. 
However, the following objects built from $\omega_{M\mA\mB}$ and $\tilde \omega_{M \pA \pB}$ are in fact scalars under generalised diffeomorphisms:
\be
\begin{split}
\Phi_{\pC \mA \mB}  \equiv \Vp^M{}_{\pC} \omega_{M \mA \mB} \,, \quad
 \tilde\Phi_{\mC \pA \pB}  \equiv \Vm^M{}_{\mC}\tilde \omega_{M \pA \pB} \,, 
 \label{definephi}
\end{split} 
\ee
\be
\begin{split}
\varphi_{\mA\mB\mC}  \equiv \Vm^M{}_{[\mA} \omega_{|M| \mB \mC]}\,, \quad
 \tilde \varphi_{\pA \pB \pC}  \equiv \Vp^M{}_{[\pA}\tilde \omega_{|M| \pB \pC]}\,. 
\end{split} 
\label{definevarphi}
\ee
It is actually only these combinations that appear in the action \eqref{bmr}.
We can in fact relate the above spin connections to projections of the double field theory covariant derivative \cite{Siegel:1993th,Siegel:1993xq,Jeon:2011cn,Hohm:2011si}.
We will calculate the scalars \eqref{definephi} and \eqref{definevarphi} explicitly for the Newton-Cartan background in appendix \ref{spinconn}. 

The constraints then take the form:
\begin{equation}
\begin{split}
2\cH &= - X^{\prime M}\Pm_{MN} X^{\prime N} + i\flathm_{\mA\mB}\psi^{\mA}\psi^{\prime \mB} 
+ iX^{\prime M}(P^{N}{}_M\omega_{N\mA\mB}\psi^{\mA}\psi^{\mB}- \Pm^{N}{}_M\tilde{\omega}_{N\pA\pB}\tilde{\psi}^{\pA}\tilde{\psi}^{\pB}) 
 \\
& -\frac{1}{4}\Pp^{MN}\omega_{M\mA\mB}\omega_{N\mC\mD}\psi^{\mA}\psi^{\mB}\psi^{\mC}\psi^{\mD} 
+\frac{1}{4}\Pm^{MN}\tilde{\omega}_{M\pA\pB}\tilde{\omega}_{N\pC\pD}\tilde{\psi}^{\pA}\tilde{\psi}^{\pB}\tilde{\psi}^{\pC}\tilde{\psi}^{\pD} 
 \\
& - \frac{1}{2}F_{\mA\mB \pC \pD}\psi^{\mA}\psi^{\mB}\tilde{\psi}^{\pC}\tilde{\psi}^{\pD} \,, 
\\
2\tilde{\cH} &= X^{\prime M} \Pp_{MN} X^{\prime N} - i\flathp_{\pA \pB}\tilde{\psi}^{\pA}\tilde{\psi}^{\prime \pB} 
+ i X^{\prime M}(\Pp^{N}{}_M\omega_{N\mA\mB}\psi^{\mA}\psi^{\mB} -\Pm^{N}{}_M\tilde{\omega}_{N\pA\pB}\tilde{\psi}^{\pA}\tilde{\psi}^{\pB})
\\
&-\frac{1}{4}\Pp^{MN}\omega_{M\mA\mB}\omega_{N\mC\mD}\psi^{\mA}\psi^{\mB}\psi^{\mC}\psi^{\mD} 
+\frac{1}{4} \Pm^{MN}\tilde{\omega}_{M\pA\pB}\tilde{\omega}_{N\pC\pD}\tilde{\psi}^{\pA}\tilde{\psi}^{\pB}\tilde{\psi}^{\pC}\tilde{\psi}^{\pD}  
\\
& - \frac{1}{2}\tilde{F}_{\pA \pB \mC\mD}\tilde{\psi}^{\pA}\tilde{\psi}^{\pB}\psi^{\mC}\psi^{\mD} \,, 
\end{split}
\label{HH}
\end{equation}
where we have ``curvatures'' with flat indices:
\be
\begin{split}
	F_{\mA\mB\pC \pD} & = 
	2 \Vm^M{}_{[\mA} \partial_M ( \tilde \Phi_{\mB]\pC\pD})  + 2 \tilde \Phi_{[\mA| \pE \pC } \tilde \Phi_{|\mB] \pD}{}^{\pE}
	+ 3 \varphi_{\mA \mB \mE} \tilde \Phi^{\mE}{}_{\pC\pD} \,,
	\\ 
	\tilde F_{\pA \pB \mC\mD} & = 
	2 \Vp^M{}_{[\pA} \partial_M (  \Phi_{\pB]\mC\mD})  + 2  \Phi_{[\pA| \mE \mC }  \Phi_{|\pB] \mD}{}^{\mE}
	+ 3 \tilde\varphi_{\pA \pB \pE} \Phi^{\pE}{}_{\mC\mD} \,.
\end{split}
\label{defineFF}
\ee
These are scalars under generalised diffeomorphisms, as they involve the quantities \eqref{definephi} and \eqref{definevarphi} which are already scalars. 
Though it is tempting to think of these four-index objects as curvatures, because they couple to the four-fermion terms in the action, strictly speaking their relationship to genuine notions of curvature is more indirect. In particular, in the geometry of double field theory the generalised Riemann tensor does not give a well-defined generalised curvature tensor \cite{Jeon:2011cn,Hohm:2011si}.
If we were to unflatten the indices on $F_{\mA\mB\pC \pD}$ and $\tilde F_{\pA \pB \mC\mD}$, we would obtain four-index generalised tensors, which however would not be invariant under the local $O(D) \times O(D)$ generalised Lorentz transformations (as $F$ and $\tilde F$ are not \cite{Blair:2013noa}) and so could not be interpreted as a generalised Riemann tensor built solely out of the generalised metric, instead involving also derivatives of the generalised vielbeins.\footnote{In line with this, what one finds in the reduction to the standard string sigma model as in \cite{Blair:2013noa} (though this result was not stated explicitly there), is that $F_{\mA\mB\pC\pD} = \frac{1}{2} R_{\pm \mA\mB\pC\pD} + \frac{1}{2} \omega_{+ i \mA \mB} \omega_{-}{}^i{}_{\pC\pD} = \tilde F_{\pC\pD \mA\mB}$, where $\omega_{\pm}$ denote torsionful spin connections, and $R_\pm$ the associated (identical) Riemann curvatures (with flat indices).}

Finally, we have:
\begin{equation}
\begin{split}
-\sqrt{2}\mathcal{Q} &= X^{\prime M}\eta_{MN} \Vm^N{}_{\mA}\psi^{\mA} 
+ \frac{i}{2}\varphi_{\mA \mB \mC} \psi^{\mA}\psi^{\mB}\psi^{\mC} 
+ \frac{i}{2} \tilde \Phi_{\mC \pA \pB} \tilde{\psi}^{\pA}\tilde{\psi}^{\pB}\psi^{\mC} \,,
\\
\sqrt{2}\tilde{\mathcal{Q}} &= X^{\prime M}\eta_{MN}\Vp^N{}_{\pA}\tilde{\psi}^{\pA} 
+ \frac{i}{2} \Phi_{\pC \mA \mB} \psi^{\mA}\psi^{\mB}\tilde{\psi}^{\pC} 
+ \frac{i}{2}\tilde\varphi_{\pA \pB \pC} \tilde{\psi}^{\pA}\tilde{\psi}^{\pB}\tilde{\psi}^{\pC} \,.
\end{split}
\label{QQ}
\end{equation}
These generate worldsheet translations and supersymmetries. This can be seen explicitly by introducing symmetry parameters $\alpha, \tilde \alpha$ (bosonic) and $\epsilon, \tilde \epsilon$ (fermionic), and defining the ``smeared'' quantities:
\be
\mathcal{H}(\alpha)  \equiv \int d\sigma \alpha(\sigma) \mathcal{H}(\sigma) \,,\quad
\tilde{\mathcal{H}}(\tilde\alpha) \equiv \int d\sigma \tilde\alpha(\sigma) \tilde{\mathcal{H}}(\sigma) \,,
\ee
\be
\mathcal{Q}(\epsilon)  \equiv \int d\sigma \epsilon(\sigma) Q(\sigma)\,,\quad
\tilde{\mathcal{Q}}(\tilde\epsilon) \equiv \int d\sigma \epsilon(\sigma) Q(\sigma)\,,
\ee
Then for any quantity $\mathcal{O}$ we define its variations as:
\be
\delta_\alpha \mathcal{O} = \{ \mathcal{H}(\alpha),\mathcal{O} \}^* \,,\quad \delta_\epsilon \mathcal{O} = \{ \mathcal{Q}(\epsilon), \mathcal{O} \}^*
\,,\quad
\delta_{\tilde\alpha} \mathcal{O} = \{ \tilde{\mathcal{H}}(\tilde\alpha),\mathcal{O} \}^* \,,\quad \delta_{\tilde\epsilon} \mathcal{O} = \{ \tilde{\mathcal{Q}}(\tilde\epsilon), \mathcal{O} \}^*\,,
\label{symmetryvars}
\ee
using the Dirac brackets:
\be
\{ X^M(\sigma) , X^N(\sigma^\prime ) \}^*  = - \eta^{MN} \theta(\sigma-\sigma^\prime)\,,
\ee
\be
\{ \psi^{\mA} (\sigma) , \psi^{\mB}(\sigma^\prime) \}^* = i\flathm^{\mA \mB} \delta(\sigma-\sigma^\prime) \,,\quad
\{ \tilde\psi^{\pA} (\sigma) , \tilde\psi^{\pB}(\sigma^\prime) \}^* = i\flathp^{\pA \pB} \delta(\sigma-\sigma^\prime) \,,
\ee
where $\partial_\sigma \theta(\sigma) = \delta(\sigma)$, $\theta(-\sigma) = - \theta(\sigma)$. Note this doubled Dirac bracket ensures that we have the standard bracket $\{ X^\mu(\sigma) , P_{\nu} (\sigma^\prime) \}^* = \{ X^\mu(\sigma), \tilde X^\prime_{\nu} (\sigma^\prime)\}^* = \delta^\mu{}_\nu \delta(\sigma-\sigma^\prime)$.

For instance, the supersymmetry variations of the constraints themselves are:
\be
\delta_\epsilon \mathcal{Q} = i \epsilon \mathcal{H} \,,\quad 
\delta_\epsilon \mathcal{H} = \frac{3}{2} \epsilon^\prime \mathcal{Q} + \frac{1}{2} \epsilon \mathcal{Q}^\prime\,,\quad
\delta_{\tilde\epsilon} \tilde{\mathcal{Q}} = i \tilde\epsilon \tilde{\mathcal{H}} \,,\quad 
\delta_{\tilde\epsilon} \tilde{\mathcal{H}} = -\frac{3}{2}\tilde \epsilon^\prime \tilde{\mathcal{Q}} - \frac{1}{2} \tilde\epsilon \tilde{\mathcal{Q}}^\prime\,.
\label{susyofQH}
\ee
Before integrating out the dual coordinates to obtain a standard Lagrangian form of the action, we want to emphasise to the reader that the action \eqref{bmr} will automatically provide the Hamiltonian form of the string action on replacing $\tilde X^{\prime}_\mu = P_\mu$, and the bosonic kinetic term with $\dot{X}^\mu P_\mu$. It is necessary just to evaluate the constraints explictly for the background we are considering.

\subsection{Integrating out the dual coordinates} 

The terms in the Lagrangian of \eqref{bmr} that involve the bosonic coordinates $X^M$ are:
\be
\begin{split}
	L_X & = \frac{1}{2} \dot{X}^M \eta_{MN} X^{\prime N} 
-\frac{1}{2} e \cH_{MN} X^{\prime M} X^{\prime N}
- \frac{1}{2} u \eta_{MN} X^{\prime M} X^{\prime N}
	+ X^{\prime M} f_M \,,
\end{split}
\ee
where the final term contains the coupling to the fermions, with
\be
f_M = - ie \Pp^N{}_M \omega_{N\mA \mB} \psi^{\mA} \psi^{\mB} + ie \Pm^N{}_M \tilde \omega_{N \pA \pB} \tilde \psi^{\pA} \tilde \psi^{\pA}
+ \frac{i\xi}{\sqrt{2}} \Vm_{M \mA} \psi^{\mA} - \frac{i\tilde \xi}{\sqrt{2}} \Vp_{M \pA} \tilde \psi^{\pA}  \,,
\label{definefM}
\ee
which is a generalised vector.

We can proceed to integrate out the dual coordinates $\tilde X_\mu$ using the same procedure as for the bosonic string.
The only difference is the appearance of terms involving the fermions in $f_M$, amounting to shifting the quantity $\mathcal{C}^\mu$ defined in \eqref{C} to $\mathcal{C}^\mu + f^\mu$.
The end result is that
the Lagrangian after this integration out is given by $L = L_X + L^\prime$,
where $L_X$ is given by:
\be
\begin{split} 
	L_X & =
	\frac{1}{2} K_{\mu\nu} \left( \frac{1}{e}D_\tau X^\mu D_\tau X^\nu  - e X^{\prime \mu} X^{\prime \nu} \right) 
	+ B_{\mu\nu}  \dot X^{ \mu} X^{\prime \nu} 
	\\ & \qquad
	+ \frac{1}{e} K_{\mu\nu}D_\tau X^\mu f^\nu + X^{\prime \mu} \uf_\mu 
	+ \frac{1}{2e} K_{\mu\nu} f^\mu f^\nu 
	\\ & \qquad
	+  \beta_a x_\mu^a ( D_-X^\mu + f^\mu)  + \bar \beta_{\bar a} \bar x_\mu^{\bar a} ( D_+ X^\mu + f^\mu) \,,
\end{split}
\label{lxhereagain}
\ee
where now $\uf_\mu = f_\mu - B_{\mu\nu} f^\nu$, $D_\tau \equiv \partial_\tau - u \partial_\sigma$, $D_\pm \equiv D_\tau \pm e \partial_\sigma$,
and the remaining solely fermionic terms are:
\be
\begin{split}
L^\prime  = &
- \frac{i}{2} \flathm_{\mA \mB} \psi^{\mA} ( \dot\psi^{\mB} - (u-e) \psi^{\prime \mB} )
- \frac{i}{2} \flathp_{\pA \pB}  \tilde \psi^{\pA} ( \dot{\tilde{\psi}}^{\pB} - (u+e) \tilde\psi^{\prime \pB} )
\\
& \qquad 
+ \frac{e}{4} \left( \flathp^{\pE \pF} \Phi_{\pE \mA\mB}\Phi_{\pF \mC\mD}\psi^{\mA}\psi^{\mB}\psi^{\mC}\psi^{\mD} + \flathm^{\mE \mF} \tilde{\Phi}_{\mE\pA\pB}\tilde{\Phi}_{\mF\pC\pD}\tilde{\psi}^{\pA}\tilde{\psi}^{\pB}\tilde{\psi}^{\pC}\tilde{\psi}^{\pD}\right)
 \\
& \qquad 
+ \frac{e-u}{4} F_{\mA \mB \pC \pD}\psi^{\mA}\psi^{\mB}\tilde{\psi}^{\pC}\tilde{\psi}^{\pD} 
+ \frac{e+u}{4} \tilde{F}_{\pA \pB \mC \mD}\tilde{\psi}^{\pA}\tilde{\psi}^{\pB}\psi^{\mC}\psi^{\mD}
 \\ 
& \qquad
+ \frac{1}{2\sqrt{2}} \xi \left(  -\varphi_{\mA\mB\mC}\psi^{\mA}\psi^{\mB}\psi^{\mC}  -\tilde{\Phi}_{\mC\pA \pB}\tilde{\psi}^{\pA}\tilde{\psi}^{\pB}\psi^{\mC} \right)
\\
& \qquad 
+ \frac{1}{2\sqrt{2}} \tilde \xi \left( \Phi_{\pC \mA\mB}\psi^{\mA}\psi^{\mB}\tilde{\psi}^{\pC} + \tilde{\varphi}_{\pA \pB\pC} \tilde{\psi}^{\pA}\tilde{\psi}^{\pB}\tilde{\psi}^{\pC} \right) \,,
\end{split}
\label{Lprime}
\ee
with the various geometric quantities here defined in \eqref{definephi}, \eqref{definevarphi} and \eqref{defineFF}.
Now, the quantities $f^\mu$ and $\uf_\mu \equiv f_{\mu} - B_{\mu\nu} f^\nu$ appearing in \eqref{lxhereagain} are the components arising from the vector $f_M$, defined in \eqref{definefM}, which we can write as
\be
f_M = - ie \Vp_M{}^{\pC} f_{\pC} - i e \Vm_M{}^{\mC} \bar f_{\mC} \,,
\label{fflat0}
\ee
with
\be
f_{\pC} \equiv  \Phi_{\pC \mA \mB} \psi^{\mA} \psi^{\mB} + \frac{\tilde\xi}{\sqrt{2} e} \tilde \psi_{\pC} \,,\quad
\bar f_{\mC} \equiv \tilde\Phi_{\mC \pA \pB} \tilde\psi^{\pA} \tilde\psi^{\pB} - \frac{\xi}{\sqrt{2} e} \psi_{\mC} 
\,.
\label{fflat1}
\ee
Using the vielbein parametrisation of section \ref{vieltech}, it follows that
\be
f^\mu = - \frac{ie}{\sqrt{2}} h^{\mu \pC} f_{\pC} - \frac{ie}{\sqrt{2}} \bh^{\mu \mC} \bar f_{\mC} \,,
\quad
\mathring{f}_{\mu} = - \frac{ie}{\sqrt{2}} k_\mu{}^{\pC} f_{\pC} + \frac{ie}{\sqrt{2}} \bar k_\mu{}^{\mC} \bar f_{\mC} \,.
\label{fflat2}
\ee
Thus far this has been completely general. The resulting Lagrangian given by the sum of \eqref{lxhereagain} and \eqref{Lprime} gives the full worldsheet supersymmetric Lagrangian for an arbitrary $(n,\bar n)$ non-Riemannian doubled background. 
Although not immediately obvious, it can be tidied up into a form which is manifestly covariant on the worldsheet and which contains the expected sort of geometric couplings to the background in the form of generalised spin connections, torsions and curvatures.
We will not present the general details of this procedure here, and instead will focus on the Newton-Cartan case, for which we will use a slightly bespoke approach to manipulating our result into an understandable form.

\subsection{Manipulations for the Newton-Cartan parametrisation} 

So, we now specialise to a Newton-Cartan parametrisation of our doubled background.
Our goal is to isolate all terms involving the additional worldsheet bosonic field $V$, so that it only appears in the constraints.  (Essentially, we want to isolate the combination $(x_\mu - \bar x_\mu) X^\mu$ which picks out the direction $V$. This is not an especially natural combination in the doubled approach, because $x_\mu$ and $\bar x_\mu$ are associated to the projectors $\Pp$ and $\Pm$ respectively, which appear everywhere.)
Let's focus on the following combination in \eqref{lxhereagain}:
\be
\begin{split} 
	+ \frac{1}{e} K_{\mu\nu}D_\tau X^\mu f^\nu + X^{\prime \mu} \uf_\mu 
	+ \frac{1}{2e} K_{\mu\nu} f^\mu f^\nu 
	+  \beta x_\mu ( D_-X^\mu + f^\mu)  + \bar \beta \bar x_\mu^{} ( D_+ X^\mu + f^\mu)
	 \,.
\end{split}
\label{Xf}
\ee
We carry out the following manipulations:
\begin{itemize}
\item We replace $D_\tau = \frac{1}{2} (D_+ + D_-)$ and $\partial_\sigma = \frac{1}{2} (D_+ -D_-)$
\item We expand $X^\mu = (X^i , V)$ and insert the explicit Newton-Cartan parametrisations of $K_{\mu\nu}$, $h^{\mu}{}_{\pC}$, $\bar h^{\mu}{}_{\mC}$, $k_{\mu}{}^{\pC}$ and $\bar k_\mu{}^{\mC}$, using \eqref{NCKHB}, \eqref{NCxybasis} and \eqref{NCviel1}, \eqref{NCviel2}. We also expand $f_{\pC} = ( f_{\um} , f_{\utau}, f_{\uu})$ and $\bar f_{\mC} = ( \bar f_{\um}, \bar f_{\utau}, \bar f_{\uu})$.
\item We note that the terms involving $\beta$ and $\bar \beta$ are:
\be
\frac{1}{\sqrt{2}} \beta \left(
D_- V + \tau_i D_- X^i - ie \sqrt{2} f_{\uu} \right)
-
\frac{1}{\sqrt{2}}\bar \beta \left(
D_+ V -\tau_i D_+ X^i + ie \sqrt{2} \bar f_{\uu} \right) \,.
\ee
\item Anywhere we have $D_\pm V$ appearing we add and subtract from it the extra terms appearing in the brackets above, so that $D_\pm V$ only appears in the combinations which are the equations of motion of $\beta$ and $\bar \beta$.
\end{itemize}
The result is that we can write \eqref{Xf} in terms of the parts that involve $V$:
\be
\begin{split}
\frac{1}{\sqrt{2}}& \left( \beta + \frac{i}{2} (f_{\utau} + \bar f_{\utau}) \right)
\left(
D_- V + \tau_i D_- X^i - ie \sqrt{2} f_{\uu} \right)
\\&
- \frac{1}{\sqrt{2}}
\left(
\bar\beta + \frac{i}{2} (f_{\utau} + \bar f_{\utau})
\right)
\left(
D_+ V -\tau_i D_+ X^i + ie \sqrt{2} \bar f_{\uu} \right)\,,
\end{split}
\label{Vhere}
\ee
and the parts that involve only $X^i$, where now we also insert the explicit expressions for the components of $f_{\pC}$ and $\bar f_{\mC}$, giving:
\be
\begin{split}
-& \frac{i}{\sqrt{2}} \left(
D_+ X^i ( e_i{}^{\um} \Phi_{\um \mA \mB} + \tau_i \Phi_{\utau \mA \mB} ) \psi^{\mA} \psi^{\mB}
+
D_- X^i ( e_i{}^{\um}\tilde\Phi_{\um \pA \pB} + \tau_i \tilde\Phi_{\utau \pA \pB} ) \tilde\psi^{\pA} \tilde\psi^{\pB}
\right) 
\\&
- \frac{i}{2e} \tilde \xi ( e_i{}^{\um} \tilde \psi_{\um} + \tau_i \tilde\psi_{\utau} ) D_+ X^i
-\frac{\tilde\xi}{2\sqrt{2}} ( \tilde\psi^{\pA} \Phi_{\pA \mB \mC}  \psi^{\mB}  \psi^{\mC} + \tilde\psi^{\pA} \delta_{\pA}^{\mD} \Phi_{\mD \pB \pC} \tilde\psi^{\pB} \tilde\psi^{\pC} )
\\ & 
+ \frac{i}{2e}  \xi ( e_i{}^{\um}  \psi_{\um} + \tau_i \psi_{\utau} ) D_- X^i
+ \frac{\xi}{2\sqrt{2}} ( \psi^{\mA} \tilde\Phi_{\mA \pB \pC} \tilde \psi^{\pB} \tilde \psi^{\pC} + \psi^{\mA} \delta_{\mA}^{\pD} \tilde\Phi_{\pD \mB \mC} \psi^{\mB} \psi^{\mC} )
\\&
- \frac{e}{4} \Big( 
\Phi_{\pE \mA \mB} \Phi^{\pE}{}_{\mC \mD}\psi^{\mA}\psi^{\mB}\psi^{\mC}\psi^{\mD} 
+\tilde\Phi_{\mE \pA \pB} \tilde\Phi^{\mE}{}_{\pC \pD}\tilde \psi^{\pA} \tilde{\psi}^{\pB}\tilde{\psi}^{\pC}\tilde{\psi}^{\pD}
\\ & \qquad\qquad\qquad \qquad\qquad\qquad\qquad+ 2 \flathp^{\pE \mF} \Phi_{\pE \mA \mB} \tilde \Phi_{\mF \pC \pD} \psi^{\mA} \psi^{\mB}\tilde{\psi}^{\pC}\tilde{\psi}^{\pD}
\Big)- \frac{1}{4e} \tilde \xi \xi \flathp_{\pA \mB} \tilde \psi^{\pA} \psi^{\mB}
\,.
\end{split}
\label{fermionsout}
\ee
Here we defined
\be
\flathp_{\pA \mB} \equiv \flathp_{\mB \pA} = \begin{pmatrix}
\delta_{\um \un} & 0 & 0 \\ 
0 & 0 & 1 \\
0 & 1 & 0 
\end{pmatrix} \,,
\ee
which captures cross-coupling between left and right projected sectors. This is numerically identical to $\flathm_{\mA \mB}$ and $\flathp_{\pA \pB}$.
It is immediately clear that there are some cancellations between \eqref{fermionsout} and \eqref{Lprime}, removing all terms involving $\psi\psi\psi\psi$, $\tilde\psi\tilde\psi\tilde\psi\tilde\psi$, $\xi \psi \tilde\psi \tilde\psi$ and $\tilde\xi \tilde\psi \psi \psi$.

\subsection{Lagrangian form of the worldsheet supersymmetric Newton-Cartan action}
\label{curvedNC}

Our result for the Newton-Cartan worldsheet supersymmetric string action can thus be written as:\footnote{We have kept everything written in one-component spinor notation: appendix \ref{ws} contains the information needed to first rewrite these as projections of two-component Majorana spinors and thus write everything in manifestly covariant worldsheet notation.}
\be
\begin{split}
S = \int d^2\sigma\, 
& 
	\frac{1}{2} h_{ij} \left( \frac{1}{e}D_\tau X^i D_\tau X^j  - e X^{\prime i} X^{\prime j} \right) 
	+ B_{\mu\nu}  \dot X^{ \mu} X^{\prime \nu} 
\\ &
- \frac{i}{2}\left(  \psi^{\mA}  \flathm_{\mA \mB} D_+ \psi^{\mB} 
+ D_+ X^i \omega_{+ i \mA \mB}\psi^{\mA} \psi^{\mB}\right)
\\ & 
- \frac{i}{2} \left( \tilde\psi^{\pA}  \flathp_{\pA \pB} D_- \tilde\psi^{\pB} 
+  D_- X^i \omega_{- i \pA \pB} \tilde\psi^{\pA} \tilde\psi^{\pB}\right)
\\ & 
- \frac{i}{2e} \tilde \xi ( e_{i\um} \tilde \psi^{\um} + \tau_i \tilde\psi^{\uu}  ) D_+ X^i
+ \frac{i}{2e}  \xi ( e_{i\um} \psi^{\um} + \tau_i \psi^{\uu} ) D_- X^i
\\ & 
- \frac{1}{12} T_{\pA \pB \pC} \tilde \xi \tilde\psi^{\pA} \tilde\psi^{\pB} \tilde\psi^{\pC} 
- \frac{1}{12} \bar T_{\mA \mB \mC}  \xi \psi^{\mA} \psi^{\mB} \psi^{\mC} 
- \frac{1}{4e} \tilde \xi \xi \flathp_{\pA \mB} \tilde \psi^{\pA} \psi^{\mB}
\\ 
& + \frac{e}{2} \mathcal{R}_{\mA \mB \pC \pD} \psi^{\mA} \psi^{\mB} \tilde\psi^{\pC} \tilde\psi^{\pD} 
\\
& + 
\frac{1}{\sqrt{2}} \left( \beta + \frac{i}{2} (f_{\utau} + \bar f_{\utau}) \right)
\left(
D_- V + \tau_i D_- X^i - ie \sqrt{2} f_{\uu} \right)
\\&
- \frac{1}{\sqrt{2}}
\left(
\bar\beta + \frac{i}{2} (f_{\utau} + \bar f_{\utau})
\right)
\left(
D_+ V -\tau_i D_+ X^i + ie \sqrt{2} \bar f_{\uu} \right)	
\,.
\end{split}
\ee
The information about the geometry is captured explicitly in the couplings to $h_{ij}$ and $B_{\mu\nu}$ (which contains the field $m_i$) and in the following quantities.
We have spin connections,
\be
\begin{split}
\omega_{+ i \mA \mB} & \equiv \sqrt{2} ( e_i{}^{\um} \Phi_{\um \mA \mB} + \tau_i \Phi_{\utau \mA \mB} ) \,,\\
\omega_{-i \pA \pB} & \equiv \sqrt{2} ( e_i{}^{\um} \tilde \Phi_{\um \pA \pB} + \tau_i \tilde \Phi_{\utau \pA \pB} ) \,,\\
\end{split} 
\label{omegasNC}
\ee
torsions,
\be
\begin{split}
T_{\pA \pB \pC} & \equiv \frac{6}{\sqrt{2}} \left( - \tilde\varphi_{\pA \pB \pC} + \delta_{[\pA }{}^{\mD} \tilde\Phi_{|\mD|\pB \pC]} \right)\,,\\
\bar T_{\mA \mB \mC} & \equiv \frac{6}{\sqrt{2}} \left( \varphi_{\mA \mB \mC} - \delta_{[\mA }{}^{\pD} \Phi_{|\pD|\mB \mC]} \right)\,,
\end{split} 
\label{torsionsNC}
\ee
and curvature
\be
\mathcal{R}_{\mA \mB \pC \pD} = \frac{1}{2} \left( F_{\mA \mB \pC \pD} + \tilde F_{ \pC \pD\mA \mB}
-2 \flathp^{\pE \mF} \Phi_{\pE \mA \mB} \tilde \Phi_{\mF \pC \pD}  \right)\,.
\label{curvNC}
\ee
In fact, it can be shown that $F_{\mA \mB \pC \pD} = \tilde F_{ \pC \pD\mA \mB}$. The cheapest way to do this is to realise that this is true in a standard Riemannian parametrisation as in \cite{Blair:2013noa} and our Newton-Cartan background can be obtained from such a background by the null duality, which does not change the value of $F$ or $\tilde F$. 

In addition, we record that
\be
\begin{split}
f_{\uu} & = \Phi_{\uu \mA \mB} \psi^{\mA} \psi^{\mB} + \frac{\tilde\xi}{\sqrt{2} e} \tilde \psi^{\utau} \,,
\quad
f_{\utau}  = \Phi_{\utau \mA \mB} \psi^{\mA} \psi^{\mB} + \frac{\tilde\xi}{\sqrt{2} e} \tilde \psi^{\uu} \,,
\\
\bar f_{\uu} & = \tilde\Phi_{\uu \pA \pB} \tilde\psi^{\pA} \tilde \psi^{\pB} - \frac{\xi}{\sqrt{2}{e}} \psi^{\utau} \,,\quad
\bar f_{\utau}  = \tilde\Phi_{\utau \pA \pB} \tilde\psi^{\pA} \tilde \psi^{\pB} - \frac{\xi}{\sqrt{2}{e}} \psi^{\uu} \,.
\end{split} 
\ee
All these quantities can be worked out explicitly in components using the results of appendix \ref{spinconn}. 
We use the parametrisation in which there is also a background $B$-field with components $\mathsf{B}_{ij}$ and field strength $\mathsf{H}_{ijk} = 3 \partial_{[i}\mathsf{B}_{jk]}$ (any components $\mathsf{B}_{iv}$ can be absorbed into a redefinition of $m_i$) and for simplicity we assume that there are no off-diagonal components of the $B$-field prior to the null dualisation, i.e. that the field $B_i$ of section \ref{addb} is zero.

Then, we find for instance that the components of the torsions \eqref{torsionsNC} turn out to be equal and to contain the contribution of the field strength of the background $B$-field:
\be
\begin{array}{cclccll}
T_{\um \un \up} & = & \mathsf{H}_{ijk} e^i{}_{\um} e^j{}_{\un} e^k{}_{\up} 
\,,&
\bar T_{\um \un \up} & = & \mathsf{H}_{ijk} e^i{}_{\um} e^j{}_{\un} e^k{}_{\up} 
\,,
\\
T_{\um \un \utau} & =& - \mathsf{H}_{ijk} e^i{}_{\um} e^j{}_{\un} v^k 
\,, &
\bar T_{\um \un \utau} & = & - \mathsf{H}_{ijk} e^i{}_{\um} e^j{}_{\un} v^k \,,
\\
T_{\um \un \uu} & = & 0
\,,&  \bar T_{\um\un\uu} &= & 0\,,
\\
T_{\um \utau \uu} & = &0 
\,,&
\bar T_{\um\utau\uu}& =& 0 \,.
\end{array}
\ee
We can also straightforwardly calculate the components of the spin connections \eqref{omegasNC}:
\be
\begin{split}
\omega_{+ i \um \un} & = e^k{}_{[m|} ( \partial_i e_{k|\un]}-\partial_k e_{i|\un]} ) - h_{ik} (\partial_j e^k{}_{[\um}) e^j_{\un]}
+ \tau_i e^j{}_{[\um} e^k{}_{\un]} \partial_{[j} m_{k]} 
- \frac{1}{2} \mathsf{H}_{ijk} e^j{}_{\um} e^k{}_{\un}\,,
\,\\
\omega_{+ i\um \utau} & = \frac{1}{2} v^k \partial_i e_{k \um} 
+ \partial_{[j} h_{k]i} e^j{}_{\um} v^k 
- e^j{}_{\um} \partial_{[i} m_{j]} + \tau_i v^j e^k{}_{\um} \partial_{[j} m_{k]}
+ \frac{1}{2} e^j{}_{\um} v^k \mathsf{H}_{ijk}
\,,\\
\omega_{+ i\um \uu} & = e^j{}_{\um} \partial_{[i} \tau_{j]}
\,,\\
\omega_{+ i\utau \uu} & = v^j \partial_{[j} \tau_{i]} \,,
\end{split} 
\label{omegares1}
\ee
and
\be
\begin{split}
\omega_{- i \um \un} & = e^k{}_{[m|} ( \partial_i e_{k|\un]}-\partial_k e_{i|\un]} ) - h_{ik} (\partial_j e^k{}_{[\um}) e^j_{\un]}
+ \tau_i e^j{}_{[\um} e^k{}_{\un]} \partial_{[j} m_{k]} 
+ \frac{1}{2} \mathsf{H}_{ijk} e^j{}_{\um} e^k{}_{\un}\,,
\,\\
\omega_{- i\um \utau} & = \frac{1}{2} v^k \partial_i e_{k \um} 
+ \partial_{[j} h_{k]i} e^j{}_{\um} v^k 
- e^j{}_{\um} \partial_{[i} m_{j]} + \tau_i v^j e^k{}_{\um} \partial_{[j} m_{k]}
- \frac{1}{2} e^j{}_{\um} v^k \mathsf{H}_{ijk}
\,,\\
\omega_{- i\um \uu} & = e^j{}_{\um} \partial_{[i} \tau_{j]}
\,,\\
\omega_{- i\utau \uu} & = v^j \partial_{[j} \tau_{i]} \,.
\end{split} 
\label{omegares2}
\ee
In fact, these are the components of the original spin connection of the background \eqref{metric1} with the null isometry, except with pieces proportional to $m_i$ removed (this is related to the redefinition of $P_\mu$ to $\tilde P_\mu$ which means that terms proportional to the bare $B$-field end up appearing multiplied by the constraints).
We can turn around the definitions \eqref{omegares1} and \eqref{omegares2} to now make sense of the scalar quantities that originally appeared in the worldsheet action, based on the results listed in from appendix \ref{spinconn}. We can write:
\be
\begin{array}{cclcccl}
\Phi_{\um \mA \mB} & = & \frac{1}{\sqrt{2}} e^i{}_{\um} \omega_{+i \mA \mB}\,, & 
\tilde \Phi_{\um \pA \pB} & = & \frac{1}{\sqrt{2}} e^i{}_{\um} \omega_{-i \pA \pB}\,, \\
\Phi_{\utau \mA \mB} & = & -\frac{1}{\sqrt{2}} v^i \omega_{+i \mA \mB}\,, & 
\tilde \Phi_{\utau \pA \pB} & = & -\frac{1}{\sqrt{2}} v^i \omega_{-i \pA \pB}\,, \\
\end{array} 
\ee
while we also have 
\be
\begin{split}
\varphi_{\um\un\up} & = \frac{1}{\sqrt{2}} e^i{}_{[\um} \omega_{|i|\un\up]} -\frac{1}{6 \sqrt{2}} \mathsf{H}_{ijk} e^i{}_{\um} e^j{}_{\un} e^k{}_{\up} \,,\\
\tilde\varphi_{\um\un\up} & = \frac{1}{\sqrt{2}} e^i{}_{[\um} \omega_{|i|\un\up]} + \frac{1}{6 \sqrt{2}} \mathsf{H}_{ijk} e^i{}_{\um} e^j{}_{\un} e^k{}_{\up} \,,\\
\end{split} 
\ee
where we let $\omega_{\pm i \um \un} = \omega_{ i \um \un} \mp \frac{1}{2} \mathsf{H}_{ijk} e^j{}_{\um} e^k{}_{\un}$,
and
\be
\begin{split}
\varphi_{\um\un\utau} & = \frac{1}{\sqrt{2}} \left(
2 e^i{}_{[\um} \omega_{|i|\un]\utau} 
- v^i \omega_{i \um \un} 
\right)+ \frac{1}{6 \sqrt{2}} \mathsf{H}_{ijk} e^i{}_{\um} e^j{}_{\un} v^k \,,\\
\tilde \varphi_{\um\un\utau} & = \frac{1}{\sqrt{2}} \left(
2 e^i{}_{[\um} \omega_{|i|\un]\utau} 
- v^i \omega_{i \um \un} 
\right)- \frac{1}{6 \sqrt{2}} \mathsf{H}_{ijk} e^i{}_{\um} e^j{}_{\un} v^k \,,\\
\end{split} 
\ee
where we let $\omega_{\pm i \um \utau} = \omega_{ i \um \utau} \pm \frac{1}{2} \mathsf{H}_{ijk} e^j{}_{\um} v^k$.
These are the only components in which $\mathsf{H}_{ijk}$ appears.
We also have the components carrying a $\uu$ index, for which:
\be
\begin{split}
\Phi_{\uu \um \un} & = \tilde \Phi_{\uu \um\un}  = -3 \varphi_{\um \un \uu } = - 3 \tilde\varphi_{\um \un \uu} = -\frac{1}{\sqrt{2}} e^i{}_{\um}e^j{}_{\un} \partial_{[i} \tau_{j]}\,,\\
\Phi_{\uu \um \utau} & = \tilde \Phi_{\uu \um\utau}  = -3 \varphi_{\um \utau \uu } = - 3 \tilde\varphi_{\um \utau \uu} = \frac{1}{\sqrt{2}} e^i{}_{\um} v^j \partial_{[i} \tau_{j]} \,,
\end{split} 
\ee
with components involving the index $\uu$ twice vanishing. 
In fact, if the Newton-Cartan background is assumed to be ``twistless'' \cite{Christensen:2013lma,Christensen:2013rfa} then 
\be
h^{ik} h^{jl} \partial_{[k} \tau_{l]} = 0 \Leftrightarrow 
\tau_{[i} \partial_{j} \tau_{k]} = 0
\Leftrightarrow
e^i{}_{\um} e^j{}_{\un} \partial_{[i} \tau_{j]} = 0 
\ee
and then we have $\Phi_{\uu \um \un} = \tilde \Phi_{\uu \um\un}  = \varphi_{\um \un \uu } =  \tilde\varphi_{\um \un \uu} =0$.

What would be interesting now to do is to take the above torsionful spin connections, which we may claim are the string's preferred connections for the Newton-Cartan geometry, and use them as the building blocks appearing not only in the curvature \eqref{curvNC} but in the action and equations of motion of double field theory. Note that the background field equations of the doubled string are the equations of motion of double field theory \cite{Berman:2007xn,Copland:2011wx}. This tells us that we can derive the field equations of a Newton-Cartan background by inserting the appropriate parametrisation of the generalised metric and generalised dilaton. The results could then then be checked against the beta functional equations derived directly from the non-relativistic worldsheet theory starting with the bosonic Newton-Cartan string \cite{Gomis:2019zyu,Gallegos:2019icg}. We defer detailed investigation of the geometry and dynamics for future work.

\subsection{Supersymmetry transformations}
\label{susytransfs}

\subsubsection*{General expressions}

Let us write down the supersymmetry transformations following from \eqref{symmetryvars}.
The general expressions are \cite{Blair:2013noa}:
\be
\begin{split}
\delta_\epsilon X^M & = 
\frac{\epsilon}{\sqrt{2}} \Vm^M{}_{\mA} \psi^{\mA}
\\ & 
\quad- \int d\sigma^\prime \frac{\epsilon(\sigma^\prime) }{\sqrt{2}} \theta(\sigma-\sigma^\prime)
\Big( \psi^{\mA} X^{\prime P} \partial^M \Vm_{P \mA}
\\ & \qquad\qquad\qquad\quad
 + \frac{i}{2} \partial^M \varphi_{\mA \mB \mC} \psi^{\mA}\psi^{\mB} \psi^{\mC} 
 + \frac{i}{2} \partial^M \tilde \Phi_{\mA \pB \pC} \psi^{\mA} \tilde\psi^{\pB} \tilde\psi^{\pC} 
 \Big)(\sigma^\prime)
\,,\\
\delta_\epsilon \psi^{\mA} & = \frac{\epsilon}{\sqrt{2}} \left(
-i X^{\prime M} \Vm_{M}{}^{ \mA}
+ \frac{3}{2}\varphi^{\mA}{}_{\mB \mC} \psi^{\mB} \psi^{\mC} 
+ \frac{1}{2} \tilde\Phi^{\mA}{}_{\pB \pC} \tilde\psi^{\pB} \tilde \psi^{\pC}  
\right) \,,\\
\delta_\epsilon \tilde \psi^{\pA} & = -\frac{\epsilon}{\sqrt{2}} \tilde \Phi_{\mC}{}^{\pA}{}_{\pB} \psi^{\mC} \tilde\psi^{\pB} \,,
\end{split}
\label{susyeps}
\ee
\be
\begin{split}
\delta_{\tilde\epsilon} X^M & = -\frac{\tilde\epsilon}{\sqrt{2}} \Vp^M{}_{\pA} \tilde\psi^{\pA}
\\ & 
\quad 
+ \int d\sigma^\prime \frac{\tilde\epsilon(\sigma^\prime) }{\sqrt{2}} \theta(\sigma-\sigma^\prime)
\Big( \tilde\psi^{\pA} X^{\prime P} \partial^M \Vp_{P \pA}
\\ & \qquad\qquad\qquad\quad
 + \frac{i}{2} \partial^M \tilde\varphi_{\pA \pB \pC} \tilde\psi^{\pA}\tilde\psi^{\pB} \tilde\psi^{\pC} 
 + \frac{i}{2} \partial^M  \Phi_{\pA \mB \mC} \tilde\psi^{\pA} \psi^{\mB} \psi^{\mC} 
 \Big)(\sigma^\prime)
\,,\\
\delta_{\tilde\epsilon} \tilde\psi^{\pA} & = \frac{{\tilde\epsilon}}{\sqrt{2}} \left(
i X^{\prime M} \Vp_M{}^{ \pA}
- \frac{3}{2}\tilde\varphi^{\pA}{}_{\pB \pC} \tilde\psi^{\pB} \tilde\psi^{\pC} 
- \frac{1}{2} \Phi^{\pA}{}_{\mB \mC} \psi^{\mB} \tilde \psi^{\mC}  
\right) \,,\\
\delta_{\tilde\epsilon}  \psi^{\mA} & = +\frac{{\tilde\epsilon}}{\sqrt{2}}  \Phi_{\pC}{}^{\mA}{}_{\mB} \tilde \psi^{\pC} \psi^{\mB} \,,
\end{split}
\label{susyteps}
\ee
while the worldsheet metric and gravitino components transform as:
\be
\begin{array}{llllll}
\delta_\epsilon \lambda & = & \xi \epsilon \,, & \delta_{\tilde\epsilon} \tilde\lambda & = & \tilde\xi \tilde\epsilon \,,\\
\delta_\epsilon \xi & = & i \left( D_+ \epsilon - \frac{1}{2} \lambda^\prime \epsilon \right) \,, & 
\delta_{\tilde\epsilon} \tilde\xi & = & i \left( D_- \tilde \epsilon + \frac{1}{2} \tilde\lambda^\prime \tilde \epsilon \right) \,.
\end{array}
\ee
The transformation of the coordinates $X^M$ involves non-local expressions. However, these only affect the transformations of the components $\tilde X_\mu$, which (assuming the background obeys the section condition and does not depend on these coordinates) only appear in the action as the derivatives $\tilde X_\mu^\prime$, and we do not see this non-locality in practice.

\subsubsection*{Supersymmetry transformations for Newton-Cartan background}

Thus far these supersymmetry expressions are entirely general and apply to any doubled RNS string action.
Now let's specialise them to the Newton-Cartan background.
The key to making use of the expressions \eqref{susyeps} and \eqref{susyteps} is to recall that we had
\be
\begin{split} 
\tilde X^\prime_\mu \equiv P_\mu &  = \tilde P_\mu + B_{\mu\nu} X^{\prime \nu}\,,\\
 & =\frac{1}{e} K_{\mu\nu} ( D_\tau X^\nu + f^\nu ) + x_\mu \beta + \bar x_\mu \bar \beta + B_{\mu\nu} X^{\prime \nu}\,,
\end{split}
\ee
as a result of integrating the dual coordinates out of the action. This can be inserted into the transformation rules to determine the transformations of the fermions $\psi^{\mA}$ and $\tilde\psi^{\pA}$ in terms of $(X^\mu, \beta, \bar\beta)$.
Meanwhile, the transformations of $\beta$ and $\bar \beta$ can be determined using the fact that our definitions imply $\beta = y^\mu ( \tilde X^\prime_{\mu} - B_{\mu\nu} X^{\prime \nu})$ and $\bar \beta = \bar y^{\mu} ( \tilde X^\prime_{\mu} - B_{\mu\nu} X^{\prime \nu})$. 
Note that we have
\be
\begin{split}
\delta_\epsilon ( \tilde X^{\prime}_{\mu} - B_{\mu\nu} X^{\prime \nu} ) 
 & = - \frac{1}{2} \bar k_{\mu \mA} ( \epsilon \psi^{\mA})^\prime 
 \\ &\qquad
  + \frac{1}{2} \epsilon \psi^{\mA}  \left( 2 \partial_{[\mu} \bar k_{\nu] \mA} - \partial_\mu\bar h^\nu{}_{\mA} ( \tilde X^\prime_\nu - B_{\nu \rho} X^{\prime \rho} ) 
  - T_{\mu\nu\rho} X^{\prime \nu} \bar h^{\rho}{}_{\mA} \right)
 \\ & \qquad
 - \frac{i}{2\sqrt{2}} \epsilon \left( \partial_\mu \varphi_{\mA \mB \mC} \psi^{\mA} \psi^{\mB} \psi^{\mC} + \partial_\mu \tilde\Phi_{\mA \pB \pC} \psi^{\mA} \tilde\psi^{\pB} \tilde\psi^{\pC}  \right) \,,
\end{split} 
\ee
\be
\begin{split}
\delta_{\tilde\epsilon} ( \tilde X^{\prime}_{\mu}  - B_{\mu\nu} X^{\prime \nu} ) 
 & = - \frac{1}{2}  k_{\mu \pA} ( \tilde \epsilon \psi^{\pA})^\prime 
 \\ &\qquad
  + \frac{1}{2} \tilde \epsilon\tilde \psi^{\pA}  \left( 2 \partial_{[\mu}  k_{\nu] \pA} + \partial_\mu h^\nu{}_{\pA} ( \tilde X^\prime_\nu - B_{\nu \rho} X^{\prime \rho} ) 
  + T_{\mu\nu\rho} X^{\prime \nu}  h^{\rho}{}_{\pA} \right)
 \\ & \qquad
 + \frac{i}{2\sqrt{2}} \tilde\epsilon \left( \partial_\mu \varphi_{\pA \pB \pC} \tilde\psi^{\pA} \tilde \psi^{\pB}  \tilde\psi^{\pC} + \partial_\mu \Phi_{\pA \mB \mC}  \tilde\psi^{\pA} \psi^{\mB} \psi^{\mC}  \right) \,,
\end{split} 
\ee
so the transformation rule for $\beta$ and $\bar\beta$ will in general be rather involved. We refrain from going into the details.

For the other transformations, we will be more explicit. We have for $X^\mu = (X^i, V)$ that
\be
\begin{split}
\delta X^i & = \frac{1}{2} \epsilon ( e^i{}_{\um} \psi^{\um} - v^i \psi^{\utau} ) - \frac{1}{2} \tilde\epsilon ( e^i{}_{\um} \tilde\psi^{\um} - v^i \psi^{\utau} )\,,\\
\delta V & = - \frac{1}{2} \epsilon \psi^{\utau} - \frac{1}{2} \tilde \epsilon \tilde \psi^{\utau} \,.
\end{split}
\label{NCsusyx}
\ee
The fermion transformations work out as:
\be
\begin{split} 
\delta \psi^{\um} 
& = 
\epsilon \left( - \frac{i}{2e}  e_i{}^{\um} D_-X^i 
+ \frac{1}{2\sqrt{2}} \left(
3\varphi^{\um}{}_{\mB \mC} 
-  \Phi^{\um}{}_{\mB \mC}  
\right) \psi^{\mB}  \psi^{\mC} 
- \frac{1}{4e} \left( \tilde\xi \tilde\psi^m - \xi \psi^m \right)
\right)
\\ & \qquad
+\frac{{\tilde\epsilon}}{\sqrt{2}}  \Phi_{\pC}{}^{\um}{}_{\mB} \tilde \psi^{\pC} \psi^{\mB} 
\,,\\
\delta \tilde \psi^{m}
 & = \tilde\epsilon\left( \frac{i}{2e} e_{i}{}^{\um} D_+X^i
+ \frac{1}{2\sqrt{2}} \left(
 -3 \tilde\varphi^{\um}{}_{\pB \pC} 
+ \tilde\Phi^{\um}{}_{\pB \pC} 
\right)\tilde\psi^{\pB} \tilde\psi^{\pC} 
+ \frac{1}{4e} ( \tilde\xi \tilde\psi^{\um} - \xi \psi^{\um} )
\right)
\\ & \qquad
 -\frac{\epsilon}{\sqrt{2}} \tilde \Phi_{\mC}{}^{\um}{}_{\pB} \psi^{\mC} \tilde\psi^{\pB}
\,,\\
\end{split} 
\label{NCsusypsi1}
\ee
and
\be
\begin{split} 
\delta \psi^{\utau} & = 
\epsilon \left(
- \frac{i}{2e} (V^{\prime}  - X^{\prime i} \tau_i )
+ \frac{1}{2\sqrt{2}} \left(
3\varphi^{\utau}{}_{\mB \mC} \psi^{\mB} \psi^{\mC} 
+  \tilde\Phi^{\utau}{}_{\pB \pC} \tilde\psi^{\pB} \tilde \psi^{\pC}  
\right) 
\right)
+\frac{{\tilde\epsilon}}{\sqrt{2}}  \Phi_{\pC}{}^{\utau}{}_{\mB} \tilde \psi^{\pC} \psi^{\mB} 
\,,\\
\delta \tilde \psi^{\utau} & = 
\tilde\epsilon \left(  +\frac{i}{2e}( V^\prime + X^{\prime i} \tau_i   )
- \frac{1}{2\sqrt{2}} \left(
 3\tilde\varphi^{\utau}{}_{\pB \pC} \tilde\psi^{\pB} \tilde\psi^{\pC} 
+  \Phi^{\utau}{}_{\mB \mC} \psi^{\mB} \tilde \psi^{\mC}  
\right)
\right)
 -\frac{\epsilon}{\sqrt{2}} \tilde \Phi_{\mC}{}^{\utau}{}_{\pB} \psi^{\mC} \tilde\psi^{\pB}
\,,
\end{split}
\label{NCsusypsi2}
\ee
and
\be
\begin{split} 
\delta \psi^{\uu} & = 
\epsilon \left( - \frac{i}{\sqrt{2}}  \bar\beta
+ \frac{1}{2\sqrt{2}} \left(
3\varphi^{\uu}{}_{\mB \mC} \psi^{\mB} \psi^{\mC} 
+  \tilde\Phi^{\uu}{}_{\pB \pC} \tilde\psi^{\pB} \tilde \psi^{\pC}  
\right) 
\right)
+\frac{{\tilde\epsilon}}{\sqrt{2}}  \Phi_{\pC}{}^{\uu}{}_{\mB} \tilde \psi^{\pC} \psi^{\mB} 
\,,\\
\delta \tilde \psi^{\uu} & = 
\tilde\epsilon
\left(+
\frac{i}{\sqrt{2}} \beta
- \frac{1}{2\sqrt{2}} \left(
 3\tilde\varphi^{\uu}{}_{\pB \pC} \tilde\psi^{\pB} \tilde\psi^{\pC} 
+  \Phi^{\uu}{}_{\mB \mC} \psi^{\mB} \tilde \psi^{\mC}  
\right)
\right)
 -\frac{\epsilon}{\sqrt{2}} \tilde \Phi_{\mC}{}^{\uu}{}_{\pB} \psi^{\mC} \tilde\psi^{\pB}
\,.
\end{split} 
\label{NCsusy3}
\ee
Note that the combinations actually appearing in the action are:
\be
\begin{split}
\boldsymbol{\beta} & \equiv \frac{1}{\sqrt{2}} \left( \beta + \frac{i}{2} \left(\Phi_{\utau \mA \mB} \psi^{\mA} \psi^{\mB} +  + \tilde\Phi_{\utau \pA \pB} \tilde\psi^{\pA} \tilde \psi^{\pB} 
+ \frac{1}{\sqrt{2} e} (  \tilde\xi\tilde \psi^{\uu} - \xi \psi^{\uu}\right) \right)\,,\\
\boldsymbol{\bar{\beta}} & \equiv \frac{1}{\sqrt{2}} \left( \bar\beta + \frac{i}{2} \left(\Phi_{\utau \mA \mB} \psi^{\mA} \psi^{\mB} +  + \tilde\Phi_{\utau \pA \pB} \tilde\psi^{\pA} \tilde \psi^{\pB} 
+ \frac{1}{\sqrt{2} e} (  \tilde\xi\tilde \psi^{\uu} - \xi \psi^{\uu}\right) \right)\,,
\end{split} 
\ee
in terms of which \eqref{NCsusy3} take a form similar to \eqref{NCsusypsi1}. 
In particular we have that the terms in the transformations involving $\psi^{\mA} \psi^{\mB}$ involve the following:
 \be
\begin{split}
\Omega_{\um \bar A \bar B} \equiv 3 \varphi_{\um \mA \mB} -\Phi_{\um \mA \mB} \,,\quad
\Omega_{\utau \bar A \bar B} \equiv 3 \varphi_{\utau \mA \mB} -\Phi_{\utau \mA \mB} \,,\quad 
\Omega_{\uu \bar A \bar B} \equiv 3 \varphi_{\uu \mA \mB} -\Phi_{\uu \mA \mB} \,,
\end{split}
\ee
which has non-zero components
\be
\begin{array}{cllccclcccl}
\Omega_{\um \un \up} & = & \sqrt{2}e^i{}_{\un} e^j{}_{\up} \partial_{[i} e_{j] \um} \,,&
\Omega_{\utau \um \un} & = &-\sqrt{2} e^i{}_{\um} e^j{}_{\un} \partial_{[i} m_{j]} \,&
\Omega_{\uu \um \un } & = &\sqrt{2} e^i{}_{\um} e^j{}_{\un} \partial_{[i} \tau_{j]} \,,\\
\Omega_{\um \un \utau} & =& \sqrt{2}  v^i e^j{}_{\un} \partial_{[i} e_{j]\um}\,,&
\Omega_{\utau \um \utau} & = &\sqrt{2} e^{i}{}_{\um} v^j \partial_{[i} m_{j]} \,&
\Omega_{\uu \um \utau} & = &\sqrt{2} v^i e^j{}_{\um} \partial_{[i} \tau_{j]} \,.
\end{array} 
\ee
The quantities $\tilde \Omega_{\um \pA \pB}$ etc. with the analogous definition end up having identical components.

\subsubsection*{Global SUSY in a flat background}

In superconformal gauge ($e=1, u=0$, $\xi=\tilde\xi=0$) in a constant background, the action simplifies to:
\be
\begin{split}
S & = \int d^2\sigma \Big(
\frac{1}{2} h_{ij} ( \dot{X}^i \dot{X}^j - X^{\prime i} X^{\prime j} ) - m_i ( \dot{X}^i V^\prime - \dot{V} X^{\prime i} ) 
\\ & \qquad\quad\qquad
+   \frac{\beta}{\sqrt{2}}( \tau_i ( \dot{X}^i  - X^{\prime i} )  + \dot{V} - V^{\prime}) + \frac{\bar \beta}{\sqrt{2}} ( \tau_i ( \dot{X}^i + X^{\prime i} )- \dot{V} - V^{\prime} )
\\ 
& \qquad\qquad\qquad
- \frac{i}{2} \left( \delta_{\underline m \underline n} \psi^{\underline m} (\dot\psi^{\underline n} + \psi^{\prime \underline n} )
 + \psi^{\underline \tau} (\dot{\psi}^{\uu} + \psi^{\prime \uu} )  + \psi^{\underline u} ( \dot\psi^{\utau} + \psi^{\prime \utau} ) \right)
\\ & \qquad\qquad\qquad
- \frac{i}{2} \left( \delta_{\underline m \underline n} \tilde\psi^{\underline m}( \dot{\tilde{\psi}}^{\underline n} - \tilde{\psi}^{\prime \underline n} )
 + \tilde\psi^{\underline \tau} ( \dot{\tilde\psi}^{\uu} - \tilde\psi^{\prime \uu}) + \tilde\psi^{\underline u} ( \dot{\tilde\psi}^{\utau} - \tilde\psi^{\prime \utau})\right)
\Big) \,.
\end{split} 
\label{actionflatNC}
\ee
We can refine our presentation by defining
\be
\Xx^i \equiv X^i + v^i\tau_j X^j \,,\quad \tau_i \Xx^i = 0 \,,
\ee
\be
\gamma \equiv V + \tau_i X^i\,,\quad 
\bar\gamma \equiv - V + \tau_i X^i \,.
\ee
In this case the (global) supersymmetry transformations are:
\be
\delta \Xx^i =  \frac{1}{2} \epsilon e^i{}_{\um} \psi^{\um}- \frac{1}{2} \tilde\epsilon  e^i{}_{\um} \tilde\psi^{\um}
\,,\quad
\delta \psi^{\um} = - \frac{i}{2} \epsilon e_i{}^{\um} ( \dot{\Xx}^i - \Xx^{\prime i} )
\,,\quad
\delta \tilde \psi^{\um} = \frac{i}{2}\tilde  \epsilon e_{i}{}^{\um} ( \dot{\Xx}^i + \Xx^{\prime i} )\,,\\
\ee
\be
\delta \gamma = - \tilde\epsilon \tilde\psi^{\utau} \,,\quad
\delta \tilde\psi^{\utau} =  \frac{i}{2}\tilde \epsilon   \gamma^\prime \,,
\ee
\be
\delta \bar\gamma = -\epsilon \psi^{\utau} \,,\quad
\delta \psi^{\utau} =  \frac{i}{2} \epsilon   \bar\gamma^\prime \,,\quad
\ee
\be
\delta \beta  = - \frac{1}{\sqrt{2}} \tilde \epsilon \tilde \psi^{\prime \uu}  \,,\quad
\delta \tilde\psi^{\uu}  = \frac{i}{\sqrt{2}} \tilde \epsilon \beta  \,,
\ee
\be
\delta \bar \beta  = - \frac{1}{\sqrt{2}}  \epsilon \psi^{\prime \uu}   \,,\quad
\delta \psi^{\uu} = -\frac{i}{\sqrt{2}} \epsilon \bar \beta  \,.
\ee
One question that would naturally occur after thinking about the bosonic action would be what is the superpartner of the ``constraints'' enforced by (in our notation) $\beta$ and $\bar\beta$. The naive expectation would be that the fermions would also have to obey a constraint obtained by the supersymmetry variation of the bosonic case.
In some sense, this is true, because for instance $\beta$ imposes that $\partial_- \gamma= 0$, and the supersymmetry variation of $\partial_-\gamma$ is $\partial_-\tilde\psi^\tau$. The equation of motion for $\tilde\psi^{\uu}$ is indeed that $\partial_-\tilde \psi^\tau = 0$. 
What is really going on however is that the bosonic constraints are really chirality conditions on certain combinations of coordinates. 
The fermions $\psi$ and $\tilde\psi$ are naturally chiral, and so no additional constraints are needed.
(Note that by working with this particular flat basis we are using, these facts are especially clear. In curved indices one would need to identify the appropriate combinations of the worldsheet fermions that become chiral together. This can be read off from \eqref{psiApsimu}.)
Ultimately what is happening (in this flat case) is that part of the usual string worldsheet action involving ordinary coordinates has been replaced by a $\beta\gamma$ system, as used in the Gomis-Ooguri non-relativistic string \cite{Gomis:2000bd}, for example.

The constraints are obviously not so simple when the background is non-constant, but we would expect that there are no further fermionic constraints (with their equations of motion sufficing). We would argue that $\beta$ and $\bar\beta$ should be viewed as replacing the degrees of freedom lost by enforcing that $\gamma$ and $\bar\gamma$ are chiral, thus overall we have the same numbers of degrees of freedom in the bosonic side and hence in the fermionic side by supersymmetry. Viewed from the point of view of the parent doubled action, there is nothing unusual at play.
Nevertheless, this is something to investigate in future work.

\subsubsection*{Supersymmetric Gomis-Ooguri}

Speaking of the Gomis-Ooguri string, it is a special case of the flat Newton-Cartan background, as noted for the bosonic situation in \cite{Harmark:2018cdl}.
Let's split $i = (0,a)$ and take $h_{ab} = \delta_{ab}$, $h_{0i} = h_{00} = 0$, $v^i =( -1,0)$, $\tau_{i} =(1,0)$.
Then $\gamma \equiv V + X^0$, $\bar\gamma\equiv -V+X^0$ and the action \eqref{actionflatNC} is (dropping the total derivative term involving $m_i$):
\be
\begin{split}
S & = \int d^2\sigma \Big(
\frac{1}{2} \delta_{ab} (\dot{X}^a \dot{X}^b - X^{\prime a} X^{\prime b} ) 
+   \frac{\beta}{\sqrt{2}}\partial_- \gamma  + \frac{\bar \beta}{\sqrt{2}} \partial_+ \bar\gamma
\\ 
& \qquad\qquad\qquad
- \frac{i}{2} \left( \delta_{\underline m \underline n} \psi^{\underline m} \partial_+ \psi^{\underline n} 
 + 2\psi^{\underline u} \partial_+ \psi^{\utau}  \right)
\\ & \qquad\qquad\qquad
- \frac{i}{2} \left( \delta_{\underline m \underline n} \tilde\psi^{\underline m}\partial_- \tilde{\psi}^{\un} 
+2 \tilde\psi^{\underline u} \partial_- {\tilde\psi}^{\utau} \right)
\Big) \,.
\end{split} 
\label{actionflatNCGO}
\ee
Here we have dropped the term involving $m_i$ as it is a total derivative. However, we know from \cite{Ko:2015rha} that we should take $m_i = ( \mu,0)$ to 
generate the additional term $-\mu \partial_+  \gamma \partial_- \bar \gamma$. Here this follows from the redefinition
\be
\beta = \beta^{\text{GO}} + \frac{1}{\sqrt{2}} m_iv^i \partial_+ \bar\gamma \,,\quad
\bar \beta = \bar\beta^{\text{GO}} + \frac{1}{\sqrt{2}} m_iv^i \partial_- \gamma \,,\quad
\ee
This is a covariant way of recalling that $\beta$ and $\bar\beta$ were obtained from the shifted momentum, $\tilde P_\mu = P_\mu - B_{\mu\nu} X^{\prime}$.
In either case, the action \ref{actionflatNCGO} then gives the supersymmetric version of the Gomis-Ooguri non-relativistic string, which was studied (with $\mu=0$) in \cite{Kim:2007pc} by treating $(\beta, \gamma)$ and $( b \equiv \tilde\psi^{\uu}, c\equiv \tilde\psi^{\utau})$ as commuting $\beta\gamma$ and anticommuting $bc$ CFTs.

\section{Discussion}  
\label{discussion} 

\subsection{Surprise?} 

The purpose of this paper was to follow the author's sense of surprise and use methods and results of the doubled approach to string theory to learn about non-relativistic strings.
Depending on your perspective, we either used a null duality in the $O(D,D)$ covariant action of \cite{Blair:2013noa} or else applied the Newton-Cartan generalised metric parametrisation of section \ref{bosonic} directly to this same action.
The result, after some tidying up, is a worldsheet locally supersymmetric Newton-Cartan string, extending the bosonic action of \cite{Harmark:2018cdl}.

Let's now discuss some highlights and drawbacks of this approach, and sketch some thoughts for future directions.

\subsubsection*{Advantages and disadvantages of our approach}

The advantages of our approach include:
\begin{itemize}
\item By starting with the doubled approach, we can easily implement the null duality. 
\item The action of \cite{Blair:2013noa} also takes care of the worldsheet fermions without additional complications.  We also obtain a nice physical interpretation of the effect of the null duality on the worldsheet fermions: after the duality the separate spacetime vielbeins in the left- and right-moving sectors that the worldsheet fermions should couple to become non-invertible, and cannot be related to each other by a Lorentz transformation. This means that the effect of the background becoming non-relativistic is related to the left- and right-moving sectors on the worldsheet becoming disconnected and ``seeing'' different target spaces. In the bosonic sector, this manifests itself as the fact that the directions $\tau_i X^i \pm V$ become chiral/anti-chiral respectively.
\item Our initial action \eqref{bmr} automatically gives the Hamiltonian form of the worldsheet supersymmetric action, on replacing $\tilde X^\prime_\mu = P_\mu$.
We also automatically know the worldsheet constraints \eqref{HH} and \eqref{QQ}, the symmetry transformations they generate and their algebra, with no need to rederive or recheck this.
\item We obtain the worldsheet couplings to background geometric quantities such as spin connections, torsions and curvatures. This tells us the string's preferred structures in a Newton-Cartan background.
\item As we mentioned, we should be able to obtain an action and equations of motion for the Newton-Cartan background by directly using our Newton-Cartan generalised metric and generalised dilaton in the double field theory action and equations of motion (which also can be respectively interpreted as a generalised Ricci scalar and tensor, respectively).
This could be analysed using the full geometric machinery of double field theory \cite{Siegel:1993th,Siegel:1993xq,Jeon:2011cn,Hohm:2011si}, as perhaps could extensions to the full type II \cite{Hohm:2011dv,Jeon:2012hp} (with Ramond-Ramond fields and fermions) or heterotic \cite{Siegel:1993th,Siegel:1993xq, Hohm:2011ex} cases. 
\item The general results can be adapted to alternative parametrisations of the generalised metric which appear to describe other variants of non-relativistic geometries \cite{Morand:2017fnv}.
\end{itemize} 
The disadvantages include:
\begin{itemize}
\item After integrating out the dual coordinates, it is necessary to reconstruct the worldsheet action in a manifestly covariant form including working out explicitly the components of the doubled spin connections and related quantities. This is not entirely trivial. It remains to compare the geometric quantities we obtain with for example the spin connections obtained from the study of non-relativistic symmetry algebras e.g. in \cite{Bergshoeff:2019pij}.
\item An alternative approach which would have bypassed this perhaps lengthy detour into doubled geometry would simply have been to start with the usual locally supersymmetric RNS string in background fields (see appendix \ref{ws}) and carry out the dualisation procedure of \cite{Harmark:2018cdl} directly there! We believe this would give the same answer for the Lagrangian form of the action.
\end{itemize} 

\subsection{Exploring Newton-Cartan backgrounds in doubled geometry}

It could be interesting to explore Newton-Cartan geometry using doubled strings as a probe, or else directly using double field theory (as mentioned above).
Let's discuss first the idea of generating non-relativistic backgrounds using duality transformations in the doubled setting.
Here we are inspired by a comment made in the conclusions of \cite{Bergshoeff:2019pij} wondering about how the nature of the usual T-duality between the fundamental string solution and that of a pp-wave changes if one considers a null duality. 
We can at least easily carry out this duality in our set-up, though we will not draw any conclusions here about whether the non-relativistic background obtained has something to do with a pp-wave.
The supergravity solution of a fundamental string is:
\be
ds^2 = H^{-1} ( -dt^2  + dz^2 ) + d \vec{y}_8{}^2 \,,\quad
B = (H^{-1} - 1) dt \wedge dz \,,\quad
e^\phi = H \,,\quad
H \equiv 1 + \frac{h}{|\vec{y}_8|^6} \,.
\label{F1}
\ee
Let $w = (t +z)/\sqrt{2}$, $u= (z-t)/\sqrt{2}$, and $x^i = ( w, \vec{y}_8)$.
Then we have a Lorentzian metric with a null isometry in $u$ (there is also a null isometry in $w$), of the form \eqref{metric1} with 
\be
h_{ij} = \begin{pmatrix}
\delta_{ab} & 0 \\
0 & 0 \\
\end{pmatrix} \,,\quad
\tau_i = \begin{pmatrix} \vec{0}\\  H^{-1} \end{pmatrix}  \,,\quad
v^i = \begin{pmatrix}\vec{0} \\ - H \end{pmatrix} \,,\quad
m_i = 0 \,,
\ee
but also with a $B$-field $B_{uw} = 1- H^{-1}$. 
The generalised metric after null duality on $u$ admits the general $(1,1)$ parametrisation with (here $\mu = ( i, v) = ( a,w,v)$ where $v$ denotes the direction dual to $u$ as before)
\be
K_{\mu\nu} = \begin{pmatrix} 
h_{ij} & 0 \\
0 & 0 
\end{pmatrix} \,,\quad
H^{\mu\nu} = \begin{pmatrix} 
h^{ij} & 0 \\
0 & 0 
\end{pmatrix} \,,\quad
B_{\mu\nu} = 0\,,
\ee
(where $h^{ab} = \delta^{ab}$ and otherwise zero) and the null vectors
\be
x_\mu = \frac{1}{\sqrt{2}}\begin{pmatrix}
\vec{0} \\ 2H^{-1} - 1 \\ 1
\end{pmatrix} \,,\quad
\bar x_\mu = \frac{1}{\sqrt{2}} \begin{pmatrix}
\vec{0} \\ 1 \\ -1 
\end{pmatrix}
\,,\quad
y^\mu = \frac{1}{\sqrt{2}} \begin{pmatrix}
\vec{0} \\ H\\  H 
\end{pmatrix} 
\,,\quad
\bar y^\mu = \frac{1}{\sqrt{2}}
\begin{pmatrix} 
\vec{0} \\ H \\ H-2
\end{pmatrix} \,.
\ee
This conforms to the parametrisation \eqref{NCKHB_b} and \eqref{NCxybasis_b} incorporating the extra covector $B_i$ arising from the mixed components $B_{iu}$ of the original $B$-field. Here $B_i = ( \vec{0},H^{-1} - 1)$. Thus the (bosonic) Newton-Cartan string action in such a background is given by \eqref{NCPolyakovB}. 

Another intriguing possibility is to study backgrounds in which the string becomes non-relativistic at a singular locus. 
The example we have in mind (based on \cite{Blair:2016xnn, Berman:2019izh}) consists of the supergravity solutions that appear to describe negative branes, for instance the negative F1 solution has the form \eqref{F1} but with $H$ replaced by $\tilde H = 1 - \frac{h}{\vec{y}_8^6}$. These can be obtained by acting with timelike dualities, for instance the Buscher rules applied on both the $t$ and $z$ directions of \eqref{F1} gives this negative F1 solution.
At the point in such a solution where $\tilde H = 0$ there is a naked spacetime singularity. However, certain brane probes do not see this singularity and as a result it has been argued that one can attempt to make some sense of them in string theory \cite{Dijkgraaf:2016lym}. 
For example, a doubled string in the negative F1 background has generalised metric and dilaton
\be
\cH_{MN} = 
\begin{pmatrix}
\tilde H - 2 & 0 & 0 & 0& \tilde H - 1 & 0 \\
0 & 2 - \tilde H & 0 & \tilde H - 1 & 0 & 0\\
0 & 0 & I_8 & 0& 0 & 0\\
0 & \tilde H-1& 0 & \tilde H &  0 &0\\
\tilde H-1 & 0 & 0  & 0 & \tilde H & 0 \\
0 & 0& 0& 0& 0 & I_8
\end{pmatrix}\,,\quad
e^{-2d} = 1 \,.
\ee
At the point $\tilde H = 0$, the bottom right block of the generalised metric is non-invertible, and the generalised metric is exactly of the type $(1,1)$ form that describes the Newton-Cartan geometry we have studied in this paper. 
This is therefore a background in which for $\tilde H > 0$ and $\tilde H < 0$ we have strings probing a relativistic geometry (however with different potentially ``exotic'' variants of string theory in each region, possibly with different signatures of spacetime \cite{Dijkgraaf:2016lym}), while at the naively singular region $\tilde H = 0$ in spacetime the string theory sees a non-relativistic background. It would be very interesting to find other examples of such behaviour, and to understand whether such backgrounds should really be taken seriously.

\subsection{Other future directions} 

We considered a worldsheet supersymmetric string; it would be interesting now to compare and perhaps generalise the work of \cite{Park:2016sbw} on the doubled Green-Schwarz string (see also \cite{Gomis:2016zur} for a non-relativistic superstring).

One direction in double field theory which would be particularly appealing to pursue is whether one can adopt the techniques of generalised Scherk-Schwarz twists to obtain deformations of the Newton-Cartan geometry. The idea here (for a review see e.g. \cite{Aldazabal:2013sca}) is to study factorisable doubled backgrounds, with $\mathcal{H}_{MN}( X,\tilde X ) = U_M{}^A(X,\tilde X) U_N{}^B(X,\tilde X) \hat{\mathcal{H}}_{AB}(X)$, where the twist matrices $U_M{}^A(X,\tilde X)$ must satisfy certain consistency conditions, including that they give rise to constant generalised fluxes $f_{ABC}$. This gives a deformed theory involving the dynamical generalised metric $\hat{\mathcal{H}}_{AB}(X)$ and these fluxes. In this setting, the section condition can be relaxed, and the twist matrices can actually depend on a coordinate dual to those that appear in $\hat{\mathcal{H}}_{AB}$. However the consistency conditions ensure that this dual coordinate dependence does not explicitly enter the action or symmetries.
It would be interesting to apply this procedure in the Newton-Cartan parametrisation (note that the mechanics of this sort of twisting has some similarities to our treatment of the extra $B$-field in section \ref{addb}). One initial suggestion would be to consider whether it is consistent to let $\tau_i$, $m_i$ or $B_i$ have a linear dependence on the null direction $U$.

We can also easily generalise the approach of this paper to the exceptional sigma model \cite{Arvanitakis:2017hwb,Arvanitakis:2018hfn} which describes a U-duality covariant string action. Here we would need to know the appropriate embedding of the Newton-Cartan geometry into the generalised metrics of the U-duality groups - some possibilities were described in \cite{Berman:2019izh}. This would presumably at least reproduce the $(p,q)$ string actions of \cite{Kluson:2019ifd}.

Our final suggestion is that it would be very interesting to continue building on \cite{Berman:2019izh} in order to describe non-relativistic M-theory geometries and thus study the non-relativistic non-perturbative duality web.

\appendix

\section*{Acknowledgements} 

I am supported by an FWO-Vlaanderen Postdoctoral Fellowship, and in part by the FWO-Vlaanderen through the project G006119N and by the Vrije Universiteit Brussel through the Strategic Research Program ``High-Energy Physics''. I would like to thank David Berman, Lorenzo Menculini, Niels Obers and Jeong-Hyuck Park for useful discussions, and am especially grateful to Gerben Oling for both useful discussions and helpful feedback on a draft of this paper.

\section{Worldsheet conventions}
\label{ws}

\subsection{Conventions}

We record here our worldsheet conventions, following our earlier paper \cite{Blair:2013noa}.
The worldsheet metric can be parametrised in terms of $\lambda$ and $\tilde \lambda$ (for $\lambda \neq - \tilde \lambda$) as:
\be
\gamma_{\alpha \beta} =  \begin{pmatrix}
	-\lambda \tilde\lambda & \frac{1}{2} ( \tilde \lambda-\lambda) \\ \frac{1}{2} ( \tilde \lambda-\lambda) & 1 
\end{pmatrix} \,,\quad
(E^{-1})^{\alpha}{}_{\bar \alpha} =  \begin{pmatrix}
\frac{2}{\lambda+\tilde\lambda} & 0 \\
\frac{\lambda-\tilde\lambda}{\lambda+\tilde\lambda} & 1
\end{pmatrix} \,.
\ee
(We could also include a conformal scale, but this drops out of the action, so we exclude it completely this appendix.)
It is convenient to define $e = \frac{1}{2}(\lambda+\tilde\lambda)$ and $u=\frac{1}{2}(\tilde\lambda-\lambda)$, in terms of which the above are
\be
\gamma_{\alpha \beta} =  \begin{pmatrix}
	u^2 - e^2 & u\\ u& 1 
\end{pmatrix} \,,\quad
(E^{-1})^{\alpha}{}_{\bar \alpha} = \begin{pmatrix}
\frac{1}{e} & 0 \\
-\frac{u}{e} & 1
\end{pmatrix} \,.
\ee
The inverse metric is
\be
\gamma^{\alpha \beta} = - \frac{1}{e^2} \begin{pmatrix} 1 & -u \\ - u & u^2 -e^2 \end{pmatrix} \,.
\ee
So for instance
\be
\sqrt{-\gamma} \gamma^{\alpha \beta} \partial X^\mu \partial_\beta X^\nu g_{\mu\nu} 
= - \frac{1}{e} g_{\mu\nu} ( \dot X^{\mu} -u X^{\prime \mu} ) ( \dot X^{\nu} - u X^{\prime \nu} ) + e g_{\mu\nu} X^{\prime \mu} X^{\prime \nu} \,.
\ee
Flat gamma matrices $\gamma^{\bar \alpha}$ can be chosen as $\gamma^{\bar 0} = i\sigma_2$, $\gamma^{\bar 1} = - \sigma_1$ with $\gamma_3 = - \sigma_3$.
Then the curved ones are
\be
\gamma^0 = \frac{1}{e} \begin{pmatrix} 0 & 1 \\ -1 & 0 \end{pmatrix}\,,\quad
\gamma^1 = \begin{pmatrix}
0 & -1 - \frac{u}{e} \\ 
-1 + \frac{u}{e} & 0 
\end{pmatrix}\,.
\ee
These obey
\be
\gamma^\alpha \gamma^\beta + \gamma^\beta \gamma^\alpha = 2 \gamma^{\alpha \beta}
\ee
For two component spinors, $\bar \chi = \chi^T \gamma^{\bar{0}}$. Then for instance
\be
\sqrt{-\gamma}
\gamma^\alpha \partial_\alpha = \begin{pmatrix}
0 & \partial_\tau - (e+u) \partial_\sigma \\ 
- \partial_\tau - (e-u) \partial_\sigma & 0 
\end{pmatrix}\,,
\ee
such that
\be
\frac{i}{2} 
\sqrt{-\gamma}
\bar\Psi^A \gamma^\alpha \partial_\alpha \Psi^B \eta_{AB} 
= \frac{i}{2} \left( 
-\psi^A ( \dot\psi^B+ (-u+e) \psi^{\prime B} )
- \tilde\psi^A ( \dot{\tilde{\psi}}^B -(u+e)  ) \psi^{\prime B}
\right) \eta_{AB}
\ee
for $\Psi^A =( \psi^A , \tilde \psi^A )$.
Also,
\be
\sqrt{-\gamma}\bar \Psi^A \gamma^\alpha \partial_\alpha X \Psi^B 
= \psi^A \psi^B ( - \dot{X} - (e-u) X^\prime ) 
+ \tilde\psi^A \tilde \psi^B ( - \dot{X} + (e+u) X^\prime )\,,
\ee
\be
\sqrt{-\gamma}\bar \Psi^A \gamma^\alpha \gamma_3\partial_\alpha X \Psi^B 
= \psi^A \psi^B ( \dot{X}+ (e-u) X^\prime ) 
+ \tilde\psi^A \tilde \psi^B ( - \dot{X} + (e+u) X^\prime )\,.
\ee
We can similarly work out the bilinears:
\be
\bar\Psi^A \Psi^B = \psi^A \tilde\psi^B + \psi^B \tilde \psi^A \,,\quad
\bar \Psi^A \gamma_3 \Psi^B = \psi^A \tilde\psi^B - \psi^B \tilde \psi^A\,.
\ee
Just as we have excluded the conformal scale from our metric parametrisation (as it drops out of the action), there are components of the gravitino which do not appear due to its transformations under Weyl and super-Weyl transformations. The latter act as  $\delta_\eta \chi_\alpha = \gamma_\alpha \eta$ (and trivially on all other fields), and it is convenient to think of having fixed these transformations such that the gravitino is gamma-traceless, that is $\gamma^\alpha \chi_\alpha = 0$. 
This imposes the conditions
\be
\tilde \chi_0 = (e+u) \tilde \chi_1 \,,\quad \chi_0 = (-e+u) \chi_1 \,,
\ee
on its components. One can calculate for instance
\be
i \sqrt{-\gamma} \bar\chi_\alpha \gamma^\beta \gamma^\alpha \Psi \partial_\beta X
= i \sqrt{-\gamma} \bar \chi_\alpha \Psi \gamma^{\alpha \beta} \partial_\beta X
= 2i \tilde \chi_1 \psi ( \dot{X} - (e+u) X^{\prime} )
 + 2i \chi_1 \tilde \psi ( \dot{X} + ( e- u)X^{\prime} ) \,.
\ee

\subsection{Usual worldsheet supersymmetric string in background fields}

\subsubsection*{The full action}

The action for an RNS string in background metric and $B$-field is (from \cite{Bergshoeff:1985qr} but here following the slightly different conventions of \cite{Blair:2013noa}):
\be
\begin{split} \label{eq:rnssigmamodel}
S = -\frac{1}{2} \int &d\tau d\sigma \, \sqrt{-\gamma}\left( \gamma^{\alpha
\beta}\partial_{\alpha}X^{\mu} \partial_{\beta}X^{\nu} g_{\mu \nu} +
\epsilon^{\alpha \beta}\partial_{\alpha}X^{\mu} \partial_{\beta}X^{\nu} B_{\mu
\nu} \right.  \\
& -i \bar\Psi^\mu\gamma^{\alpha}\partial_{\alpha}\Psi^{\nu}g_{\mu \nu} - i\bar{\Psi}^{\mu}\gamma^{\alpha}\Psi^{\rho}\Gamma_{\sigma \rho}{}^{\nu}\partial_{\alpha}X^{\sigma}g_{\mu \nu}  -\frac{i}{2}\bar{\Psi}^{\mu}\gamma^{\alpha}\gamma_3\Psi^{\nu}\partial_{\alpha}X^{\rho}T_{\mu\nu\rho}  \\
&+\frac{1}{6}R_{\mu\rho\nu\sigma}\bar{\Psi}^{\mu}\Psi^{\nu}\bar{\Psi}^{\rho}\Psi^{\sigma} + \frac{1}{8}\nabla_{\rho}T_{\mu\sigma\nu}\bar{\Psi}^{\mu}\Psi^{\rho}\bar{\Psi}^{\nu}\gamma_3\Psi^{\sigma}   -\frac{1}{16}T_{\mu\rho\kappa}T^{\kappa}\,_{\nu\sigma}\bar{\Psi}^{\mu}\gamma_3\Psi^{\rho}\bar{\Psi}^{\nu}\gamma_3\Psi^{\sigma} 
\\ & -2i\bar{\chi}_{\alpha}\gamma^{\beta}\gamma^{\alpha}\Psi^{\mu}\partial_{\beta}X^{\nu}g_{\mu\nu}   \left. - \frac{1}{6}\bar{\chi}_{\alpha}\gamma^{\beta}\gamma^{\alpha}\Psi^{\mu}\bar{\Psi}^{\nu}\gamma_{\beta}\gamma_3\Psi^{\rho}T_{\mu\nu\rho} + \frac{1}{2}\bar{\chi}_{\alpha}\gamma^{\beta}\gamma^{\alpha}\chi_{\beta}\bar{\Psi}^{\mu}\Psi^{\nu}g_{\mu\nu}\right)\,.
\end{split}
\ee
Here $\Psi^\mu$ are two-component worldsheet Majorana spinors, and $\chi_\alpha$ is the worldsheet gravitino. 
We denote the field strength of the $B$-field by $T_{\mu \nu \rho} = 3 \partial_{[\mu} B_{\nu \rho]}$, the usual Levi-Civita connection by $\Gamma_{\nu \rho}{}^\mu$ and define the Riemann tensor by $R^\mu{}_{\nu \rho \sigma} = 2 \partial_{[\rho} \Gamma_{\sigma]\nu}{}^\mu +2 \Gamma_{[\rho| \lambda}{}^\mu \Gamma_{\sigma] \nu}{}^\lambda$.
In components, we write the spinor $\Psi^\mu$ as $\Psi^\mu = ( \psi^\mu, \tilde\psi^\mu)$.
We can introduce a spacetime vielbein $e_\mu{}^A$, so that $G_{\mu \nu} = e_\mu{}^A e_\nu{}^B \boldsymbol{h}_{AB}$ where $\boldsymbol{h}_{AB}$ is the flat Minkowski metric. Using this, we flatten the spacetime indices on the worldsheet fermions, $\psi^A = e_{\mu}{}^A \psi^\mu$, $\tilde \psi^A = e_\mu{}^A \tilde\psi^\mu$. In fact, one could flatten the different Weyl components by separate vielbein $e_\mu{}^A$ and $\bar e_\mu{}^{\bar A}$. It is always possible to do this after the fact, by following the indices.

\subsubsection*{Fermion kinetic terms} 

These are
\be
L_f = \frac{i}{2} \sqrt{-\gamma} \left(
\bar\Psi^\mu \gamma^\alpha \partial_\alpha \Psi^\nu g_{\mu\nu} 
+ \bar \Psi^\mu \gamma^\alpha \Psi^\nu \partial_\alpha X^\rho \Gamma_{\rho \nu}{}^\sigma g_{\mu \sigma} 
+ \frac{1}{2} \bar \Psi^\mu \gamma^\alpha \gamma_3 \Psi^\nu \partial_\alpha X^\rho T_{\mu\nu\rho}
\right)   \,,
\ee
With $D_\tau \equiv \partial_\tau - u \partial_\sigma$, $D_\pm = D_\tau \pm e \partial_\sigma$ and
\be
\Gamma_{\pm \mu \nu}{}^\rho = \Gamma_{\mu\nu}{}^\rho \pm \frac{1}{2} T_{\mu}{}^\rho{}_\nu \,,
\label{gammapm}
\ee
we have
\be
L_f =- \frac{i}{2} \left(
\psi^\mu \left[ D_+ \psi^\nu + D_+ X^\rho \Gamma_{-\rho \sigma}{}^\nu \psi^\sigma \right] g_{\mu\nu} 
+ 
\tilde \psi^\mu \left[ D_- \psi^\nu + D_- X^\rho \Gamma_{+\rho \sigma}{}^\nu \tilde\psi^\sigma \right] g_{\mu\nu} 
\right) \,.
\ee
If we then define $\psi^\mu = e^\mu{}_A \psi^A$ and
\be
\omega_{\pm \mu}{}^A{}_B = e_{\nu}{}^A \partial_\mu e^{\nu}{}_B + \Gamma_{\pm \mu \nu}{}^\rho e_{\rho}{}^A e^{\nu}{}_B \,,
\ee
we have
\be
L_f =- \frac{i}{2} \left(
\psi^A  D_+ \psi^A \flathp_{AB} 
+ \psi^A \psi^B D_+ X^\rho \omega_{-\rho AB} 
+ 
\tilde\psi^A  D_- \tilde\psi^B \flathp_{AB} 
+ \tilde\psi^A \tilde\psi^B D_- X^\rho \omega_{+\rho AB} 
\right) \,.
\ee

\subsubsection*{Gravitino terms}

These are
\be
L_\chi = \sqrt{-\gamma}\left(
i\bar{\chi}_{\alpha}\gamma^{\beta}\gamma^{\alpha}\Psi^{\mu}\partial_{\beta}X^{\nu}g_{\mu\nu}  + \frac{1}{12}\bar{\chi}_{\alpha}\gamma^{\beta}\gamma^{\alpha}\Psi^{\mu}\bar{\Psi}^{\nu}\gamma_{\beta}\gamma_3\Psi^{\rho}T_{\mu\nu\rho} - \frac{1}{4}\bar{\chi}_{\alpha}\gamma^{\beta}\gamma^{\alpha}\chi_{\beta}\bar{\Psi}^{\mu}\Psi^{\nu}g_{\mu\nu}\right)\,.
\ee
One finds
\be
\begin{split}
L_\chi & = 2 i \tilde \chi_1 \psi^\mu D_- X^\nu g_{\mu \nu} 
+ 2 i \chi_1 \tilde\psi^\mu D_+ X^\nu g_{\mu\nu}
\\ & \qquad
+ \frac{e}{3} T_{\mu\nu\rho} \left( 
\chi_1 \tilde\psi^\mu \tilde \psi^\nu \tilde \psi^\rho
- \tilde \chi_1 \psi^\mu \psi^\nu \psi^\rho
\right) 
- 4 e \chi_1 \tilde \chi_1 \psi^\mu \tilde \psi^\nu g_{\mu\nu} \,.
\end{split} 
\ee
The identification used in \cite{Blair:2013noa} and in the main body of the present paper is then: 
\be
\tilde \chi_1 = \frac{\xi}{4e} \,,\quad \chi_1 = - \frac{\tilde \xi}{4e}\,.
\ee

\subsubsection*{Four-fermion terms}

These are
\be
\begin{split}
L_{\psi\psi\tilde\psi\tilde\psi} & = \sqrt{-\gamma} \Big( -\frac{1}{12}R_{\mu\rho\nu\sigma}\bar{\Psi}^{\mu}\Psi^{\nu}\bar{\Psi}^{\rho}\Psi^{\sigma} - \frac{1}{16}\nabla_{\rho}T_{\mu\sigma\nu}\bar{\Psi}^{\mu}\Psi^{\rho}\bar{\Psi}^{\nu}\gamma_3\Psi^{\sigma} \\ & \qquad\qquad\qquad\qquad\qquad\qquad\qquad\qquad  + \frac{1}{32}T_{\mu\rho\kappa}T^{\kappa}\,_{\nu\sigma}\bar{\Psi}^{\mu}\gamma_3\Psi^{\rho}\bar{\Psi}^{\nu}\gamma_3\Psi^{\sigma} \Big)\\
 & = \frac{e}{4} R_{\pm \mu\nu\rho\sigma} \psi^{\mu} \psi^{\nu} \tilde\psi^{\rho}\tilde\psi^{\sigma} \,,
\end{split} 
\ee
where $R_{+\mu\nu\rho\sigma} = R_{-\mu\nu\rho\sigma}$ are the Riemann tensors for the torsionful connections \eqref{gammapm} defined above. These are equal by the Bianchi identity for the three-form field strength.

\section{Details of the spin connections}
\label{spinconn} 

\subsubsection*{The spin connections in terms of doubled vielbeins}

We consider
\be
\omega_{M \mA \mB} = - \Vm_{N \mA} \partial_M \Vm^N{}_{\mB} + \Vm^N{}_{[\mA} \Vm^P{}_{\mB]} \partial_P \cH_{MN} \,,\quad
\tilde\omega_{M \pA \pB} = \Vp_{N \pA} \partial_M \Vp^N{}_{\pB} + \Vp^N{}_{[\pA} \Vp^P{}_{\pB]} \partial_P \cH_{MN} \,.
\ee
Writing $\cH_{MN} = \Vm_{M \mA} \Vm_N{}^{\mA} + \Vp_{M \pA} \Vp_N{}^{\pA}$ we obtain the equivalent forms:
\be
\begin{split}
\omega_{M\mA\mB} & = - \Vm_{N \mA} \partial_M \Vm^N{}_{\mB} - \Vm^P{}_{[\mA|} \partial_P \Vm_{M|\mB]} 
 + \Vm_M{}^{\mC} \Vm^N{}_{[\mA} \Vm^P{}_{\mB]} \partial_{P} \Vm_{N \mC} 
 + \Vp_M{}^{\pC} \Vm^N{}_{[\mA} \Vm^P{}_{\mB]} \partial_{P} \Vp_{N \pC} \,,
\\ 
\tilde\omega_{M\pA\pB} & =  +\Vp_{N \pA} \partial_M \Vp^N{}_{\pB} - \Vp^P{}_{[\pA|} \partial_P \Vm_{M|\pB]} 
 + \Vp_M{}^{\pC} \Vp^N{}_{[\pA} \Vp^P{}_{\pB]} \partial_{P} \Vp_{N \pC} 
 + \Vm_M{}^{\mC} \Vp^N{}_{[\pA} \Vp^P{}_{\pB]} \partial_{P} \Vm_{N \mC}  \,.
\end{split}
\ee
We define the generalised diffeomorphism scalars:
\be
\begin{split}
\Phi_{\pC \mA \mB} & \equiv \Vp^M{}_{\pC} \omega_{M \mA \mB}  = - \Vp^M{}_{\pC} \Vm_{N \mA} \partial_M \Vm^{N}{}_{\mB} - 2 \Vm^M{}_{[\mA} \Vm^N{}_{\mB]} \partial_M \Vp_{N \pC}\,,\\
 \tilde\Phi_{\mC \pA \pB} & \equiv \Vm^M{}_{\mC}\tilde \omega_{M \pA \pB} = + \Vm^M{}_{\mC} \Vp_{N \pA} \partial_M \Vp^{N}{}_{\pB} + 2 \Vp^M{}_{[\pA} \Vp^N{}_{\pB]} \partial_M \Vm_{N \mC}\,,
\end{split} 
\ee
and
\be
\begin{split}
\varphi_{\mA\mB\mC} & \equiv \Vm^M{}_{[\mA} \omega_{|M| \mB \mC]}  = - \Vm^M{}_{[\mA} \Vm^N{}_{\mB} \partial_{|M|} \Vm{}_{|N|\mC]} \,,\\
 \tilde \varphi_{\pA \pB \pC} & \equiv \Vp^M{}_{[\pA}\tilde \omega_{|M| \pB \pC]}  = + \Vp^M{}_{[\pA} \Vp^N{}_{\pB} \partial_{|M|} \Vp{}_{|N|\pC]} \,.
\end{split} 
\ee

\subsubsection*{The spin connections in terms of the non-Riemannian parametrisation}

We insert the parametrisation
\be
\Vm_{M \mA} = \frac{1}{\sqrt{2}} \begin{pmatrix} - \bar k_{\mu \mA} + B_{\mu\nu} \bar h^{\nu}{}_{\mA} \\ \bar h^{\mu}{}_{\mA} \end{pmatrix} 
\,,\quad
\Vp_{M \pA} = \frac{1}{\sqrt{2}} \begin{pmatrix}  k_{\mu \pA} + B_{\mu\nu}  h^{\nu}{}_{\pA} \\  h^{\mu}{}_{\pA} \end{pmatrix} \,.
\ee
We must have that $k_{\mu}{}_{\pA} h^{\mu}{}_{\pB} + k_{\mu}{}_{\pB} h^{\mu}{}_{\pA} = \flathp_{\pA \pB}$ and assume further that we choose a parametrisation as in section \ref{vieltech} such that $k_{\mu}{}_{\pA} h^{\mu}{}_{B}$ is constant. Then, with $T_{\mu\nu\rho} \equiv 3 \partial_{[\mu} B_{\nu\rho]}$, we have
\be
\begin{split}
\Phi_{\pC \mA \mB} & =
\frac{1}{\sqrt{2}}\left(
 h^{\mu}{}_{\pC} \bh^{\nu}{}_{[\mA|} \partial_\mu \bar k_{\nu |\mB]}
+ \bh^{\mu}{}_{[\mA|} \bar k_{\nu}{}_{|\mB]} \partial_\mu h^{\nu}{}_{\pC}
- \bh^{\mu}{}_{[\mA} \bh^{\nu}{}_{\mB]} \partial_\mu k_{\nu \pC}
\right) 
\\ & \qquad
 - \frac{1}{2\sqrt{2}} \bh^{\mu}{}_{\mA} \bh^{\nu}{}_{\mB} h^\rho{}_{\pC} T_{\mu\nu\rho}\,,\\
\tilde\Phi_{\mC \pA \pB} & =
\frac{1}{\sqrt{2}}\left(
 \bh^{\mu}{}_{\mC} h^{\nu}{}_{[\pA|} \partial_\mu  k_{\nu |\pB]}
+ h^{\mu}{}_{[\pA|}  k_{\nu}{}_{|\pB]} \partial_\mu \bh^{\nu}{}_{\mC}
- h^{\mu}{}_{[\pA} h^{\nu}{}_{\mB]} \partial_\mu \bar k_{\nu \mC}
\right) 
\\ & \qquad
 + \frac{1}{2\sqrt{2}} h^{\mu}{}_{\pA} h^{\nu}{}_{\pB} \bh^\rho{}_{\mC} T_{\mu\nu\rho}\,,
\end{split}
\ee
and
\be
\begin{split} 
\varphi_{\mA\mB\mC} & = 
\frac{1}{\sqrt{2}}
\bh^{\mu}{}_{[\mA} \bh^\nu{}_{\mB} \partial_{|\mu} \bar k_{\nu| \mC]} 
- \frac{1}{6\sqrt{2}} \bh^{\mu}{}_{[\mA} \bh^\nu{}_{\mB} \bh^{\rho}{}_{\mC]} T_{\mu\nu\rho} \,,
\\
\tilde \varphi_{\pA \pB \pC} & = 
\frac{1}{\sqrt{2}}
h^{\mu}{}_{[\pA} h^\nu{}_{\pB} \partial_{|\mu} k_{\nu| \pC]} 
+ \frac{1}{6\sqrt{2}} h^{\mu}{}_{[\pA} h^\nu{}_{\pB} h^{\rho}{}_{\pC]} T_{\mu\nu\rho} \,.
\end{split}  
\ee

\subsubsection*{The spin connections in terms of the Newton-Cartan parametrisation} 

We use the Newton-Cartan parametrisation of \eqref{NCKHB} and \eqref{NCxybasis} with
\be
h^{\mu}{}_{\pA} = \begin{pmatrix} e^{i}{}_{\um} & - v^i & 0 \\ 0 & 1 & 0 \end{pmatrix} \,,\quad
\bar h^{\mu}{}_{\mA} = \begin{pmatrix} e^{i}{}_{\um} & - v^i & 0 \\ 0 & -1 & 0 \end{pmatrix} \,,
\ee
\be
k_\mu{}^{\pA} = \begin{pmatrix} e_i{}^{\um} & \tau_i & 0 \\ 0 & 1 & 0 \end{pmatrix} \,,\quad
\bar k_\mu{}^{\mA} = \begin{pmatrix} e_i{}^{\um} & \tau_i & 0 \\ 0 &- 1 & 0 \end{pmatrix} \,.
\ee
The coordinates are $X^\mu = ( X^i, V)$, and we assume we only depend on the $X^i$.
We may write both flat indices as $\pA = ( \um, \utau, \uu)$ and $\mA = (\um, \utau, \uu)$: the distinction between them will be of relevance only on the worldsheet where they are (automatically) carried by fermions of different chirality. Then for instance we have that $h^{i}{}_{\um}$, $h^{i}{}_{\utau}$ and $k_{i}{}^{\um}$, $k_i{}^{\utau}$ are non-constant, similarly $\bh^{i}{}_{\um}$, $\bh^{i}{}_{\utau}$ and $\bar k_{i}{}^{\um}$, $\bar k_i{}^{\utau}$. 
Note that the flat metrics are taken in this parametrisation to be off-diagonal in the $(\utau,\uu)$ components, thus $k_{\mu \uu} = k_\mu{}^{\utau}$ and so on. 
We will also easily incorporate the possibility of a background $B$-field with components												
\be
B_{\mu\nu} = 
\begin{pmatrix} 
\mathsf{B}_{ij} & - m_i \\
m_j & 0 
\end{pmatrix} \,,
\ee
containing both the covector field $m_i$ and an additional contribution $\mathsf{B}_{ij}$. (Any additional piece $\mathsf{B}_{iv}$ could simply be absorbed into a redefinition of $m_i$. Note also that we are excluding the extra field $B_i \equiv B_{i u}$ that could arise from the null dualisation of a Lorentzian background with background $B$-field, i.e. we take $B_i = 0$. This is a simplifying assumption and could straightforwardly be relaxed.)
We have $T_{ijk} = \mathsf{H}_{ijk}$, $T_{ijv} = - 2 \partial_{[i} m_{j]}$ where $\mathsf{H}_{ijk} = 3 \partial_{[i} \mathsf{B}_{jk]}$.

We can then calculate the components of the generalised diffeomorphism scalars defined above. 
The end result is:
\be
\begin{split}
 \Phi_{\up \um \un} & = \frac{1}{\sqrt{2}} 
\left(
e^i{}_{\up} e^j{}_{[\um|} \partial_i e_{j|\un]} 
+ e^i{}_{[\um} e_{|j| \un]} \partial_i e^{j}{}_{\up}
- e^{i}{}_{[\um} e^j{}_{\un]} \partial_i e_{j \up} 
\right)
- \frac{1}{2\sqrt{2}} e^i{}_{\um} e^j{}_{\un} e^k{}_{\up} \mathsf{H}_{ijk} 
\,,\\
 \Phi_{\up \um \utau} & =
\frac{1}{\sqrt{2}} \left(
 e^{i}{}_{\up} v^j \partial_{[i} e_{j] \um} 
+  e^{i}{}_{\um} v^j \partial_{[i} e_{j] \up} 
-  e^i{}_{\up} e^j{}_{\um} \partial_{[i} m_{j]}
\right)
+ \frac{1}{2\sqrt{2}} e^i{}_{\up} e^j{}_{\um}  \mathsf{H}_{ijk} v^k
\,,\\
 \Phi_{\up \um \uu} & =
\frac{1}{\sqrt{2}}  e^i{}_{\up} e^j{}_{\um} \partial_{[i} \tau_{j]} 
\,,\\
 \Phi_{\up \utau \uu } & =
\frac{1}{\sqrt{2}} v^i e^j{}_{\up} \partial_{[i} \tau_{j]}
\,,\\
 \Phi_{\utau \um \un} & = 
\frac{1}{\sqrt{2}} 
\left(
v^j e^i{}_{[\um|} (\partial_i e_{j |\un]}   - \partial_j e_{i|\un]})
+ e^i{}_{\um} e^j{}_{\un} \partial_{[i} m_{j]}
\right)
+ \frac{1}{2\sqrt{2}}e^{i}{}_{\um} e^j{}_{\un}  \mathsf{H}_{ijk} v^k 
\,,\\
\Phi_{\utau \um \utau}  & = 
\frac{1}{\sqrt{2}}e^i{}_{\um} v^j ( - 2 \partial_{[i} m_{j]} ) 
\,,\\
 \Phi_{\utau \um \uu} & =
\frac{1}{\sqrt{2}} v^j e^i{}_{\um} \partial_{[i} \tau_{j]} 
\,,\\
 \Phi_{\uu  \um \un } & =
-\frac{1}{\sqrt{2}} e^i{}_{\um} e^j{}_{\un} \partial_{[i} \tau_{j]}
\,,\\
 \Phi_{\uu \um \utau } & =
\frac{1}{\sqrt{2}} e^i{}_{\um} v^j \partial_{[i}\tau_{j]}
\,,\\
\end{split}
\ee
with $ \Phi_{\uu \um \uu} =  \Phi_{\uu \utau \uu} =  \Phi_{\utau \utau \uu} =0$,
\be
\begin{split}
\tilde \Phi_{\up \um \un} & = \frac{1}{\sqrt{2}} 
\left(
e^i{}_{\up} e^j{}_{[\um|} \partial_i e_{j|\un]} 
+ e^i{}_{[\um} e_{|j| \un]} \partial_i e^{j}{}_{\up}
- e^{i}{}_{[\um} e^j{}_{\un]} \partial_i e_{j \up} 
\right)
+ \frac{1}{2\sqrt{2}} e^i{}_{\um} e^j{}_{\un} e^k{}_{\up} \mathsf{H}_{ijk} 
\,,\\
\tilde \Phi_{\up \um \utau} & =
\frac{1}{\sqrt{2}} \left(
 e^{i}{}_{\up} v^j \partial_{[i} e_{j] \um} 
+  e^{i}{}_{\um} v^j \partial_{[i} e_{j] \up} 
- e^i{}_{\up} e^j{}_{\um} \partial_{[i} m_{j]}
\right)
- \frac{1}{2\sqrt{2}} e^i{}_{\up} e^j{}_{\um}  \mathsf{H}_{ijk} v^k
\,,\\
\tilde \Phi_{\up \um \uu} & =
\frac{1}{\sqrt{2}}  e^i{}_{\up} e^j{}_{\um} \partial_{[i} \tau_{j]} 
\,,\\
\tilde \Phi_{\up \utau \uu } & =
\frac{1}{\sqrt{2}} v^i e^j{}_{\up} \partial_{[i} \tau_{j]}
\,,\\
\tilde \Phi_{\utau \um \un} & = 
\frac{1}{\sqrt{2}} \left( 
v^j e^i{}_{[\um|} (\partial_i e_{j |\un]}   - \partial_j e_{i|\un]})
+ e^i{}_{\um} e^j{}_{\un} \partial_{[i} m_{j]}
\right) 
- \frac{1}{2\sqrt{2}}e^{i}{}_{\um} e^j{}_{\un} \mathsf{H}_{ijk} v^k 
\,,\\
\tilde \Phi_{\utau \um \utau}  & = 
\frac{1}{\sqrt{2}}e^i{}_{\um} v^j ( - 2 \partial_{[i} m_{j]}) 
\,,\\
\tilde \Phi_{\utau \um \uu} & =
\frac{1}{\sqrt{2}}
  e^i{}_{\um} v^j \partial_{[i} \tau_{j]} 
\,,\\
\tilde \Phi_{\uu  \um \un } & =
-\frac{1}{\sqrt{2}} e^i{}_{\um} e^j{}_{\un} \partial_{[i} \tau_{j]}
\,,\\
\tilde \Phi_{\uu \um \utau } & =
\frac{1}{\sqrt{2}} e^i{}_{\um} v^j \partial_{[i}\tau_{j]}
\,,\\
\end{split}
\ee
with $\tilde \Phi_{\uu \um \uu} = \tilde \Phi_{\uu \utau \uu} = \tilde \Phi_{\utau \utau \uu} =0$,
and
\be
\begin{split}
\varphi_{\um \un \up} & = 
\frac{1}{\sqrt{2}} e^i{}_{[\um} e^j{}_{\un|} \partial_i e_{j | \up]}
-\frac{1}{6\sqrt{2}} e^i{}_{\um} e^j{}_{\un} e^k{}_{\un} \mathsf{H}_{ijk}
\,,
\,\\
\varphi_{\um \un \utau} & = 
\frac{1}{6\sqrt{2}} \left( 2 e^{i}{}_{[\um|} v^j ( \partial_{i} e_{j|\un]} - \partial_j e_{i |\un]} )
- 2 e^i{}_{\um} e^j{}_{\un} \partial_{[i} m_{j]}\right)
+ \frac{1}{6\sqrt{2}}e^i{}_{\um} e^j{}_{\un} \mathsf{H}_{ijk} v^k 
\,,\\
\varphi_{\um \un \uu} & =
\frac{1}{6\sqrt{2}} 2 e^i{}_{\um} e^j{}_{\un} \partial_{[i} \tau_{j]}
 \,,\\
\varphi_{\um \utau \uu} & =
\frac{1}{6\sqrt{2}} 2 v^i e^j{}_{\um}  \partial_{[i} \tau_{j]}
 \,,
\end{split}
\ee
\be
\begin{split}
\tilde\varphi_{\um \un \up} & = 
\frac{1}{\sqrt{2}} e^i{}_{[\um} e^j{}_{\un|} \partial_i e_{j | \up]}
+ \frac{1}{6\sqrt{2}} e^i{}_{\um} e^j{}_{\un} e^k{}_{\un} \mathsf{H}_{ijk}
\,,
\,\\
\tilde\varphi_{\um \un \utau} & =
\frac{1}{6\sqrt{2}} \left( 2 e^{i}{}_{[\um|} v^j ( \partial_{i} e_{j|\un]} - \partial_j e_{i |\un]} )
- 2 e^i{}_{\um} e^j{}_{\un} \partial_{[i} m_{j]}\right)
- \frac{1}{6\sqrt{2}} e^i{}_{\um} e^j{}_{\un}\mathsf{H}_{ijk} v^k
 \,,\\
\tilde\varphi_{\um \un \uu} & = 
\frac{1}{6\sqrt{2}} 2 e^i{}_{\um} e^j{}_{\un} \partial_{[i} \tau_{j]}
\,,\\
\tilde\varphi_{\um \utau \uu} & = 
\frac{1}{6\sqrt{2}} 2 v^i e^j{}_{\um}  \partial_{[i} \tau_{j]}
\,,
\end{split}
\ee
Note that the components of $\Phi$ and $\tilde \Phi$, and $\varphi$ and $\tilde\varphi$, are all equal except for the terms involving $H_{ijk}$, which change sign.

\bibliography{NewBib}

\end{document}